\documentclass[a4paper,11pt]{article}
\usepackage[english]{babel}
\usepackage{jheppub}
\usepackage{subfig}
\usepackage[T1]{fontenc} % if needed
\usepackage{graphicx}
\usepackage{epsfig}
\usepackage{axodraw}
\usepackage{amsmath}
\usepackage{amssymb}
\usepackage{float}
\usepackage{placeins}

\allowdisplaybreaks[4]

\title{Planar master integrals for the two-loop light-fermion electroweak corrections 
to Higgs plus jet production}

\author[a,b]{Matteo Becchetti}
\author[a,b]{Roberto Bonciani}
\author[a,b]{Valerio Casconi}
\author[c]{Vittorio Del Duca\footnote{On leave from INFN, Laboratori Nazionali di Frascati, Italy.} }
\author[c]{Francesco Moriello}

\affiliation[a]{Sapienza - Universit\`a di Roma, Dipartimento di Fisica, Piazzale Aldo Moro 5, 00185, Rome, Italy}

\affiliation[b]{INFN Sezione di Roma, Piazzale Aldo Moro 2, 00185, Rome, Italy}

\affiliation[c]{ETH Z\"{u}rich, Institut f\"{u}r theoretische Physik, Wolfgang-Pauli
str. 27, 8093, Z\"{u}rich, Switzerland}

\emailAdd{matteo.becchetti@roma1.infn.it}
\emailAdd{roberto.bonciani@roma1.infn.it}
\emailAdd{valerio.casconi@roma1.infn.it}
\emailAdd{delducav@itp.phys.ethz.ch}
\emailAdd{fmoriello@phys.ethz.ch}

\abstract{We present the analytic calculation of the planar master integrals which contribute to compute the two-loop light-fermion electroweak corrections to the production of a Higgs boson in association with a jet in gluon-gluon fusion. The complete dependence
on the electroweak-boson mass is retained. The master integrals are evaluated by means of the differential equations method and the analytic results are expressed in terms of multiple
polylogarithms up to weight four.}

\newcommand{\be}{\begin{equation}}
\newcommand{\ee}{\end{equation}}
\newcommand{\nn}{\nonumber}
\newcommand{\bea}{\begin{eqnarray}}
\newcommand{\eea}{\end{eqnarray}}
\newcommand{\bfig}{\begin{figure}}
\newcommand{\efig}{\end{figure}}
\newcommand{\bc}{\begin{center}}
\newcommand{\ec}{\end{center}}

\newcommand{\T}{\mathcal{T}}
\newcommand{\D}{\mathcal{D}}
\newcommand{\intd}{\int\,\D^{d}k_{1}\D^{d}k_{2}\,}
\newcommand{\Space}{\qquad\qquad\qquad}

\begin{document}

\maketitle
\flushbottom

\section{Introduction}

Since its discovery in 2012~\cite{Aad:2012tfa,Chatrchyan:2012ufa}, the study of the Higgs boson has been one of the dominant topics in the physics programme of the Large Hadron Collider (LHC).
The Higgs signals provide a strong test of the Standard Model (SM) and can be a probe
of New Physics (NP), in case a deviation from the SM behaviour of the Higgs properties 
will be revealed~\cite{Dittmaier:2011ti,Dittmaier:2012vm,Heinemeyer:2013tqa,deFlorian:2016spz}.

The major production channel of the Higgs boson at the LHC is gluon-gluon fusion. 
Since the Higgs boson does not couple directly to gluons, the coupling is mediated by a heavy-quark loop.
Thus, leading-order production requires the evaluation of one-loop amplitudes, production at NLO accuracy requires the evaluation of two-loop amplitudes, 
and so forth. So far, in the full theory, {\it i.e.} with a complete dependence on the heavy-quark mass, Higgs production via gluon-gluon fusion
is known at NLO~\cite{Graudenz:1992pv,Spira:1995rr} (see also~\cite{Harlander:2005rq,Anastasiou:2006hc,Aglietti:2006tp,Bonciani:2007ex}).  
However, in the limit in which the heavy-quark mass is much heavier than the other scales in the process, 
the computations are made more tractable
by using the Higgs Effective Field Theory (HEFT), i.e. by replacing the loop-mediated Higgs-gluon coupling by a tree-level effective coupling.
That lowers by one loop the amplitudes to be computed.
In the HEFT, Higgs production via gluon-gluon fusion is known at ${\rm N^3LO}$~\cite{Anastasiou:2013srw,Anastasiou:2013mca,Anastasiou:2014lda,Li:2014afw,Anastasiou:2015ema}, whose accuracy has reached the 5\% level~\cite{Anastasiou:2016cez}. 

In order to have access to a detailed study of the Higgs properties, precise theoretical predictions for more exclusive observables are needed. 
Of particular importance, for a detailed study of the SM structure of the loop-mediated Higgs-gluon coupling and of possible NP contributions to it,
is the prediction of the transverse momentum distribution of a Higgs boson at high $p_T$~\cite{Harlander:2013oja,Banfi:2013yoa,Azatov:2013xha,Grojean:2013nya,Schlaffer:2014osa,Buschmann:2014twa,Dawson:2014ora,Buschmann:2014sia,Ghosh:2014wxa,Dawson:2015gka,Langenegger:2015lra,Azatov:2016xik,Grazzini:2016paz}.

%particular, one of the most interesting one is to compare the Higgs couplings with respect to the SM predictions. Indeed, despite the fact that new particles could be in principle at energy scales not yet available, the effects of the physics beyond the standard model (BSM) can be revealed in deviation of the Higgs couplings from the SM predictions.

%In this context, gluon-fusion can be used to investigate several features of the Higgs physics: top Yukawa coupling, coupling to gluons, Composite Higgs scenarios, etc. In the SM Lagrangain there is no interaction vertex among gluons and Higgs boson, therefore the gluon-fusion process is mediated by loop diagrams, which are almost entirely dominated by top massive loops, already at the Leading Order (LO).

%The Higgs production by gluon-fusion can be studied in the context of the Effective Field Theories (EFTs), in this way it is possible to modify the SM Langrangian with higher dimensional operator which allow for point-like interaction between gluons and Higgs boson: as an example, we refer to \cite{Azatov:2013xha, Azatov:2016xik} where the modified Lagrangian depends on two parameters, namely $c_t$ and $c_g$, which parametrize the top-Higgs coupling and the effective-point-like coupling to gluons. Despite the fact that the total cross section is proportional to the sum of the parameter $c_t$ and $c_g$, it is possible to study certain exclusive process that are sensitive to the single couplings.

The transverse momentum distribution of the Higgs boson in gluon-gluon fusion is known in the HEFT at NNLO~\cite{Boughezal:2013uia,Chen:2014gva,Boughezal:2015dra,Boughezal:2015aha,Chen:2016zka}. 
While this approximation is satisfactory in the region in which the Higgs $p_T$ is smaller than the top-quark mass, in the high-$p_T$ region 
it is known to give rise to predictions that can differ considerably from the ones in which the top-quark mass is treated exactly~\cite{Grazzini:2013mca}.
In the full theory, the production of a Higgs boson in association with a jet is known only at leading order~\cite{Ellis:1987xu,Kauffman:1991jt}.
Several approximations exist to Higgs plus one jet production at NLO in the full theory: 
$(i)$ a numerical computation~\cite{Jones:2018hbb},
which includes, though, only the dependence on the top-quark mass, while neglecting the bottom-quark mass;
$(ii)$ computations in the high $p_T$ region~\cite{Lindert:2018iug,Neumann:2018bsx}, based on
the two-loop amplitudes for Higgs plus three partons, computed in the limit when the Higgs transverse momentum is larger than the top-quark mass,
$p_T\gg m_t$~\cite{Kudashkin:2017skd}.;
$(iii)$ computations in the intermediate $p_T$ region, where it is relevant to include 
also the top-bottom interference~\cite{Lindert:2017pky,Caola:2018zye}, which is estimated by interfering a top-quark loop computed in the HEFT 
with a bottom-quark loop computed as an expansion in a small bottom-quark mass~\cite{Melnikov:2016qoc,Melnikov:2017pgf}.

The exact evaluation of the Higgs plus one jet production at NLO in the full theory, {\it i.e.} retaining the complete dependence on the mass of the heavy quark that runs into the loops, requires to compute the two-loop four-point amplitudes for Higgs plus three partons, where one of the loops is a heavy-quark loop,
which involve elliptic iterated integrals. So far, only the planar contributions have been computed~\cite{Bonciani:2016qxi}.

In addition, although the Higgs boson does not couple directly to massless fermions, its coupling to light fermions may be mediated by an electroweak-boson loop. In particular, the Higgs boson may also couple to gluons via a double loop, made by an electroweak-boson loop and a light-quark loop~\cite{Aglietti:2004nj,Aglietti:2004ki}. This production mode makes up the bulk of the two-loop electroweak corrections to Higgs production
via gluon-gluon fusion~\cite{Degrassi:2004mx,Degrassi:2005mc,Actis:2008ug} and increases by about 5\% the leading-order gluon-fusion cross section
and by about 2\% the ${\rm N^3LO}$ cross section~\cite{Anastasiou:2016cez}.

The NLO QCD corrections to the two-loop light-quark contribution to Higgs production via gluon-gluon fusion consist of the three-loop mixed QCD-electroweak
corrections to Higgs production~\cite{Bonetti:2016brm,Bonetti:2017ovy}, and of the two-loop four-point amplitudes for Higgs plus three partons, where one loop is an electroweak-boson loop and the other is a light-quark loop. These two-loop four-point amplitudes are presently unknown.
Two approximated evaluations of the NLO QCD corrections have been computed: $(a)$ the full corrections in the un-physical limit where
the Higgs mass is much smaller than the electroweak-boson mass~\cite{Anastasiou:2008tj}; $(b)$ a computation in which the virtual part is treated 
exactly~\cite{Bonetti:2017ovy}, but the real radiation is included in the soft-gluon approximation~\cite{Bonetti:2018ukf}.

%In the kinematic region in which the transverse momentum of the Higgs boson is bigger
%(or much bigger) than the top mass, electroweak corrections could play a non negligible role, and they should be taken into account in the study of the process. In spite of the fact that the perturbative order of these corrections is formally suppressed with respect to the pure QCD channels, the behavior of logarithmic terms 
%in the ratio of the large scale over the mass of the W boson could have an impact, 
%that has to be evaluated. Among the electroweak corrections, a leading role is played
%by the corrections due to the Feynman diagrams at the two-loop level that contain a 
%closed loop of light fermions coupled to the Higgs via W exchange
%This kind of corrections occur at the two-loop level for the first time and, therefore, are UV finite and free from final state IR singularities, if the transverse momentum of the Higgs is different from zero. They are of order ${\mathcal O}(g_S^3 g_w^3)$ in the amplitude and, therefore, give rise to corrections starting at ${\mathcal O}(\alpha_S^3 \alpha_w^2)$ in the cross section, when interfered with the leading order. 
%
%When integrated over the additional parton in the final state, the light-fermion
%corrections provide the $2 \to 2$ contribution to the production of a resonant Higgs,
%considered in 

In this paper, we present the analytic computation of the planar master integrals (MIs) relevant to the evaluation of the 
two-loop four-point amplitudes for Higgs plus three gluons, where one loop is an electroweak-boson loop and the other is a light-quark loop.
They contribute to the mixed QCD-electroweak light-fermion corrections to the production of a Higgs boson with an additional jet,
as well as to the real radiation of the NLO QCD corrections to the two-loop light-quark contribution to Higgs production via gluon-gluon fusion.

%These MIs are, therefore, needed to the evaluation of the EW NNLO planar corrections to the full inclusive Higgs production and to the NLO EW corrections to Higgs production associated with one jet. \par
%As is well known, the main production channel of the Higgs boson at the LHC is the gluon-fusion one. In the Standard Model (SM) Lagrangain there is no interaction vertex among gluons and Higgs boson, therefore the gluon-fusion process is mediated by loop diagrams already at the Leading Order (LO), this makes the computation of radiative corrections non-trivial. This loop interaction is almost entirely dominated by top massive loops, as a consequence QCD effects are the main source of the radiative corrections to this process. In this direction, the fully inclusive Higgs production cross section is known up to NLO \cite{Graudenz:1992pv, Spira:1995rr}, while the cross section of the Higgs production in association with one jet \cite{Ellis:1987xu} and the Higgs $p_T$ distribution \cite{Kauffman:1991jt} are known at the LO. \par
%%Regarding the QCD corrections to Higgs plus one jet production, recently, progress has been made. Indeed, all the planar two-loop MIs relevant for the comptutation of the two-loop scattering amplitude for the process $H\rightarrow 3$ partons with full dependence on top mass have been computed \cite{Bonciani:2016qxi}. On the other hand, the NLO QCD corrections to the Higgs $p_T$ distribtion have been computed in the large transverse momentum, $p_t \gg 2m_t$, limit \cite{Kudashkin:2017skd, Lindert:2018iug}. \par
%
We performed the reduction to the MIs using the computer softwares \texttt{FIRE5} \cite{Smirnov:2014hma} and \texttt{LiteRed} \cite{Lee:2013mka}\footnote{Other public available softwares \cite{Anastasiou:2004vj, vonManteuffel:2012np, Maierhoefer:2017hyi} for the IBP reduction exist.}, which implement IBPs \cite{Chetyrkin:1981qh, Tkachov:1981wb, Laporta:2001dd} and Lorentz-invariance \cite{Gehrmann:1999as} identities, obtaining 48 MIs. A part from the two-point functions, these MIs are new and presented in this paper for the first time. They are analitically computed using the differential equations method \cite{Kotikov:1990kg, Remiddi:1997ny, Gehrmann:2002zr, Argeri:2007up, Henn:2014qga}. This method has proved to be very efficient for the computation of the MIs needed for higher-order corrections in the SM. In particular, we adopt the canonical basis approach \cite{Henn:2013pwa, Henn:2014qga, Argeri:2014qva, Lee:2014ioa, Georgoudis:2016wff, Gehrmann:2014bfa, Becchetti:2017abb} to the solution of the system of differential equations, which is expressed in terms of Chen-iterated integrals \cite{Chen:1977oja} represented as Goncharov multiple polylogarithms \cite{Goncharov:polylog, Goncharov2007, Remiddi:1999ew} (GPLs) up to weight four. 
The solutions are evaluated numerically using the software \texttt{GiNaC} \cite{Bauer:2000cp,Vollinga:2004sn} and tested numerically against the software \texttt{FIESTA} \cite{Smirnov:2015mct} in the Euclidean and Minkowski regions of the phase space. We find agreement in both regions.

The analytic results presented in this paper are given as ancillary files uploaded 
with the arXiv submission\footnote{The ancillary files are in text format and the results, expressed as GPLs, are written in GiNac format.}. 

The paper is structured as follows. In the Section \ref{Sec2} we give our notations and we describe the kinematic fo the processes studied. In Section \ref{Sec3} we describe briefly the method of differential equations and the canonical basis approach, moreover we give the alphabet associated to the solution. In Section \ref{Sec4} we present the canonical basis and the transformation among the MIs in canonical form and the pre-canonical ones. Finally, in Section \ref{Sec5} we present our conclusion.

\section{Notations}
\label{Sec2}

In this paper we consider the partonic processes $gg \rightarrow Hg$, $gq\rightarrow qH$, $q\bar{q} \rightarrow gH$, and the crossed channels $H\rightarrow ggg$ and $H\rightarrow q\bar{q}g$. The external momenta corresponding either to gluons or to quarks are on their mass-shell $p_i^2=0$, while the external Higgs momentum $p_4^2$ is regarded as a variable. Therefore, we appoach the solution of the master integrals (MIs) of the topology as a three-scale problem, where, apart from the Higgs momentum $p_4^2$, the other two variables are the Mandelstam variables
\begin{equation}
s=\left(p_{1}+p_{2}\right)^{2},\quad
t=\left(p_{1}+p_{3}\right)^{2}.
\end{equation}
For later convenience we define the dimensionless variabiles $x,\,y,\,z$ such that
\begin{equation}
\label{Var}
x=-\frac{s}{4 m_B^2},\quad
y=-\frac{t}{4 m_B^2},\quad
z=-\frac{p_4^2}{4 m_B^2} \, ,
\end{equation}
where $m_B^2$ is the squared mass of the internal Electroweak boson ($W$ or $Z$).\par
The physical phase-space region of the kinematic invariants \eqref{Var} is
\be
x < 0, \;\; y < 0, \;\; z < x + y \, ,
\ee
for the decay channel, while for the production channel is 
\be
x < z < 0, \;\; y > 0, \;\; z > x + y \, .
\ee

The planar corrections can be computed considering the 7-denominator topology
shown in Fig.~\ref{fig1}.
%%%%%%%%%%%%%%%%%%%%%%%%%%%%%%%%%%%%%%%%%%%%%%%%%%%%%%%%%%%%%%%%%%%%%%%%
\begin{figure}
\bc
\[ 
\vcenter{
\hbox{
  \begin{picture}(0,0)(0,0)
\SetScale{1.0}
  \SetWidth{1.0}
\Line(-50,-30)(0,-30)
%\Line(30,-30)(50,-30)
\Line(-50,30)(50,30)
\Line(-30,30)(-30,-30)
\Line(30,30)(30,-30)
\Line(30,0)(0,-30)
  \SetWidth{3.5}
\DashLine(30,-30)(50,-30){3}
\Line(0,-30)(30,-30)
\Line(30,0)(30,-30)
%
%\Text(0,-40)[c]{$(3)$}
%
\end{picture}}
}
\]
\vspace*{5mm}
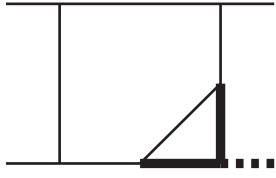
\caption{\label{fig1} Planar seven-denominator topology.}
\ec
\efig
%%%%%%%%%%%%%%%%%%%%%%%%%%%%%%%%%%%%%%%%%%%%%%%%%%%%%%%%%%%%%%%%%%%%%%%%
%
The MIs of the topology are defined by the two-loop dimensionally regularized integrals
\begin{equation}
\int \mathcal{D}^dk_1\mathcal{D}^dk_2\frac{D_{8}^{a_{8}}\,D_{9}^{a_{9}}}{D_{1}^{a_{1}}D_{2}^{a_{2}}D_{3}^{a_{3}}D_{4}^{a_{4}}D_{5}^{a_{5}}D_{6}^{a_{6}}D_{7}^{a_{7}}} \, ,
\label{integral}
\end{equation}
where $d=4-2 \epsilon$,  $a_i$ with $i = 1, \dots,7$ are integer numbers, while $a_8$ and $a_9$ are natural numbers and the normalization is such that
\be
{\mathcal D}^dk_i = \frac{d^d k_i}{i \pi^{\frac{d}{2}}} e^{\epsilon\gamma_E} \left( \frac{m_B^2}{\mu^2} \right) ^{\epsilon} \, .
\ee
The $D_i$, $i = 1,\dots,7$, are the denominators involved, while $D_8$, $D_9$ the numerators. They belong to the following set:
\begin{align}
\label{DefTop}
\big\{&-k_{1}^{2},-\left(k_{1}-k_{2} \right)^{2},-\left(p_{1}+k_{1}\right)^{2},-\left(p_{1}+p_{2}+k_{1} \right)^{2},-\left(p_{3}+k_{1}-k_{2}\right)^{2},\nonumber\\
&-\left(p_{1}+p_{2}+k_{2}\right)^{2} + m_B^2,-\left(p_{3}-k_{2} \right)^{2} + m_B^2,-\left( p_{3}-k_{1}\right)^{2},-\left( k_{2}+p_{2}\right)^{2}
\big\} \, .
\end{align}

The number of MIs for this topology is 48, considering the different channels and 
crossings. Apart from two-point functions\footnote{The two-point functions can be found for instance in \cite{Aglietti:2003yc,Aglietti:2004tq}.}, the MIs presented in this paper are new.

\section{The System of Differential Equations}
\label{Sec3}

The analytic computation of the master integrals is performed by using the differential equations method \cite{Gehrmann:2002zr,Kotikov:1990kg,Remiddi:1997ny,Argeri:2007up} applied to a canonical basis for the MIs \cite{Henn:2013pwa, Henn:2014qga}. In order to find the canonical basis several approaches exist \cite{Henn:2013pwa, Henn:2014qga, Argeri:2014qva, Lee:2014ioa, Georgoudis:2016wff}. We adopt the semi-algorithmic approach described in \cite{Gehrmann:2014bfa, Becchetti:2017abb}. The canonical basis $\vec{f}(\vec{x},\epsilon)$ satisfies a system of first order partial linear differential equations with respect to the kinematic invariants $\vec{x}$, 
\begin{equation}
\label{EqDiffPartial}
\frac{\partial\vec{f}(\vec{x},\epsilon)}{\partial x_i}= \epsilon \,A_i(\vec{x}) \vec{f}(\vec{x},\epsilon)
\end{equation}
where $i\in\{1,\dots,n\}$, $n$ is the number kinematic invariants and $A_i(\vec{x})$ is the set of matrices defining the differential equations. The linear system of partial differential equations, eq.~\eqref{EqDiffPartial}, is equivalent to the following differential form, 
\begin{equation}
\label{EqDiffFundamental}
d\vec{f}(\vec{x},\epsilon)=\epsilon\, d\tilde{A}(\vec{x})\vec{f}(\vec{x},\epsilon) \, ,
\end{equation}
where the matrix elements of $\tilde{A}(\vec{x})$ are $\mathbb{Q}$-linear combinations of logarithms, with
\begin{equation}
\frac{\partial \tilde{A}(\vec{x})}{\partial x_i}=A_i(\vec{x})\,.
\end{equation}
The set of linearly indipendent arguments of the logarithms is called the \emph{alphabet} of the solution, and its elements are called \emph{letters}.
The matrices $A_i(\vec{x})$ satisfy the integrability condition
\begin{equation}
\frac{\partial A_{j}}{\partial x_i}-\frac{\partial A_{i}}{\partial x_j}=0,\quad \left[A_{i},A_{j}\right]=0 \, ,
\end{equation}
where $\left[ A_{i}, A_{j} \right]=A_{i}A_{j}-A_{j}A_{i}$.
The solution of the differential equations \eqref{EqDiffFundamental} can be formally written as a path ordered exponential:
\begin{equation}
\label{ChenInt}
\vec{f}(\vec{x},\epsilon)=\mathbb{P}\,exp\left(\epsilon\int_{\gamma}d\tilde{A}(\vec{x}) \right)\vec{f}(\vec{x}_0,\epsilon) \, ,
\end{equation} 
where $\mathbb{P}$ stands for the path-ordering operator, $\gamma$ is a path in the space of kinematic invariants and $\vec{f}(\vec{x}_0,\epsilon)$ is a vector of boundary conditions. In practice we are interested in a solution around $\epsilon=0$. By series expanding $\vec{f}(\vec{x},\epsilon)$,
\begin{equation}
\vec{f}(\vec{x},\epsilon)=\sum_{k=0}^\infty \vec{f}^{(k)}(\vec{x}),
\end{equation}
and by parametrizing the integration contour with $t\in[0,1]$, the solution, eq.~\eqref{ChenInt}, translates to iterated integrals~\cite{Chen:1977oja}:
\begin{equation}
    \label{SolutionIteInt}
    \vec{f}(x) = \vec{f}^{(0)}(\vec{x}_0) + \sum_{k=1}^\infty\epsilon^k \sum_{j=1}^k \int_0^1 dt_1 \frac{\partial\tilde{A}(t_1)}{\partial t_1} \int_0^{t_1} dt_2 \frac{\partial\tilde{A}(t_2)}{\partial t_2} \ldots \int_0^{t_{j-1}} dt_j \frac{\partial\tilde{A}(t_j)}{\partial t_j} \,\vec{f}^{(k-j)}(\vec{x}_0)\,.
\end{equation}

In general, the alphabet letters depend algebraically on the kinematic invariants $x_i$. However, for the problem under consideration, it is possible to perform a variable change such that the alphabet depends only rationally on the new variables. This implies that the solution eq.~\eqref{SolutionIteInt} can be directly expressed in terms of Goncharov's multiple polylogarithms (GPL)~\cite{Goncharov:polylog, Goncharov2001}, defined recursively as,
\begin{equation}
 G(\alpha_{1},\dots,\alpha_{n};z)=
 \int_{0}^{z}\frac{dt}{t-\alpha_{1}} G(\alpha_{2},\dots,\alpha_{n};t) \,.
 \end{equation}  
The recursion ends when $n=0$ where we conventionally set 
\begin{equation}
G(;\,z) \equiv 1\, .
\end{equation}
Moreover, in order to deal with the divergency at the basepoint 0 when $a_n = 0$, one defines:
\begin{align}
    \label{eq:G0def}
    G(\vec{0};\,z) \equiv \frac{1}{n!}\log(z)^n\,.
\end{align}
We remark that also when the alphabet letters are not rational functions it is often possible to find a representation of the solution in terms of polylogarithmic functions. One starts from the symbol~\cite{Goncharov:2010jf} of the solution, which is obtained from the differential equations matrix $\tilde{A}(\vec{x})$ and the $\epsilon^0$ order of the boundary conditions, by the following recursive formula,
\begin{equation}
\mathcal{S}( f^{(i)}_n(\vec{x})) = \sum_m \mathcal{S}( f^{(i-1)}_m(\vec{x})) \otimes \mathcal{S}(\tilde{A}_{nm}(\vec{x}))\,.
\end{equation}
The corresponding polylogarithmic expressions are found by using the algorithm of~\cite{Goncharov:2010jf,Duhr:2011zq}, and its algebraic generalisation~\cite{Bonciani:2016qxi}.

\subsection{The Alphabet}

The system of differential equations depends originally on rational functions of $x$ and $y$ and on the following square root
%
%As already mentioned, for the process under consideration it is possible to find a change of variables such that the matrix $\tilde{A}(\vec{x})$ is a rational function of kinematical invariants. Indeed, with respect to the variables \eqref{Var} the system of differential equations depends on the root
\be
\label{root}
\sqrt{z(1+z)} \, .
\ee
Exploiting for instance the methods described in \cite{Becchetti:2017abb}, it is possible to rationalize the square root \eqref{root} by means of the change of variables
\be
\label{W}
z \rightarrow \frac{w^2}{1+2w} \, .
\ee
It is then straightforward to define the differential equations with respect to $x,y,w$ defined by the matrices $A_x(x,y,w),A_y(x,y,w),A_w(x,y,w)$ respectively and solve them by using the following iterative formula,

\begin{align}
 f^{(i)}_j(x,y,w)=\sum_k  & \left[\int_{x_0,y_0,w_0}^{x,y_0,w_0} A_{x,jk}(x,y,w)  f_k^{(i-1)}(x,y,w) dx\right.
 \nn \\
 &+ \int_{x,y_0,w_0}^{x,y,w_0} A_{y,jk}(x,y,w)  f_k^{(i-1)}(x,y,w) dy\nn \\
 &+\left.\int_{x,y,w_0}^{x,y,w} A_{w,jk} (x,y,w) f_k^{(i-1)} (x,y,w) dw\right]\nn \\
 &+f_j^{(i)}(x_0,y_0,w_0),
\end{align}
where we denoted by, e.g., $A_{x,jk}(x,y,w)$ the matrix element of the matrix $A_x(x,y,w)$ and  $x_0,y_0,w_0$ is a set of boundary points that in general depend on the master integral (see next section). The recursion above can be directly solved in terms of GPLs by factorizing the matrix elements with respect to the integration variable. By performing the factorization we obtain the (inverse) GPLs integration kernels, $(x-x_k)$,$(y-y_k)$,$(w-w_k)$, with
\bea
x_{k}&\in &\left\{0,-\frac{1}{4},\frac{1}{4}\right\} \, , \\
y_{k}&\in &\left\{0,-\frac{1}{4},\frac{1}{4},-x,-\frac{x}{4 x+1}\right\} \, , \\
w_{k}&\in &\Big\{ 0,\,-\frac{1}{2},\,-1,\,\frac{1}{4} \left(-1-i \sqrt{3}\right),\,\frac{1}{4} \left(-1+i
   \sqrt{3}\right),\,2 x,\,-\frac{2 x}{4 x+1},\,x-\sqrt{x^2+x}, \nn\\
&& x+\sqrt{x^2+x}, \frac{1}{4} \left(-1+4x\,-\sqrt{16 x^2+8 x-3}\right),
\,\frac{1}{4} \left(-1+4x+\sqrt{16 x^2+8 x-3}\right), \nn\\
&& 2 y,\,-\frac{2 y}{4 y+1},\,y-\sqrt{y^2+y}, y+\sqrt{y^2+y},\,\frac{1}{4} \left(-1+4y-\sqrt{16 y^2+8 y-3}\right), \nn\\
&& \frac{1}{4} \left(-1+4y+\sqrt{16y^2+8 y-3}\right), x+y-\sqrt{x^2+2 x y+x+y^2+y}, \nn\\
&& x+y+\sqrt{x^2+2 x y+x+y^2+y},\,\frac{-x-y}{2y},\,\frac{-x-y}{2 x} \Big\} \, ,
\eea
that correspond to the arguments of the GPLs of the solution.

\section{The Master Integrals \label{Sec4}}

The set of 48 MIs, which we refer as $\mathcal{T}_i$, $i\in\{1,\dots,48\}$, is shown in Fig.~\ref{figmasterprecan}. The corresponding canonical basis elements $f_i(\vec{x})$ are defined as linear combinations of pre-canonical integrals with algebraic prefactors of the Mandelstam invariants and the space-time regulator. Their definition is provided in appendix \ref{app:CanBasis}.

The boundary conditions for the canonical master integrals $f_3,\ldots,f_6,f_8,\ldots,f_{24}$,\newline
$f_{26},\ldots,f_{36},f_{38},f_{40},\ldots,f_{48}$ are fixed in the point $s=t=p_4^2=0$, which is a regular point for the previous canonical master integrals, instead, the master integrals $f_1,f_2,f_7,f_{25}$ are divergent in $s=t=p_4^2=0$, but they are product of known one-loop master integrals. Finally, for the master integrals $f_{37},f_{39}$ we fix the boundary conditions, respectively, in the regular points $(s = m_B^2,t=0,p_4^2=0)$ and $(s = 0, t = m_B^2, p_4^2 = 0)$.

Since we are able to express all the canonical integrals $f_{1}\dots f_{48}$ in terms of GPLs up to weight 4, the numerical
check in all the regions of the phase space is straightforward. For the numerical evaluation of the masters
we use the software \texttt{GiNaC} \cite{Bauer:2000cp,Vollinga:2004sn}. The analytic continuation in the 
Minkowski region is perfomed (numerically) adding a small imaginary part to the squared c.m. energy, 
$s+i0^+$. We checked different points of the phase space against \texttt{FIESTA} \cite{Smirnov:2015mct}, 
finding complete agreement.

%%%%%%%%%%%%%%%%%%%%%%%%%%%%%%%%%%%%%%%%%%%%%%%%%%%%%%%%%
\bfig
\bc
%\vspace*{-0.5cm}
\[
\vcenter{
\hbox{
  \begin{picture}(0,0)(0,0)
\SetScale{0.4}
  \SetWidth{1.0}
\CArc(0,0)(20,0,180)
\CArc(0,0)(20,180,360)
\CCirc(0,20){5}{0.9}{0.9}
\CCirc(34.5,23.5){5}{0.9}{0.9}
  \SetWidth{3.5}
\DashLine(-35,0)(-20,0){3}
\DashLine(20,0)(35,0){3}
\CArc(60,0)(40,150,180)
\CArc(0,34.6)(40,300,330)
\CArc(30.20,17.60)(5.28,-34,153)
\Text(-16,8)[c]{\footnotesize{$p_4^2$}}
\Text(0,-20)[c]{\footnotesize{(${\mathcal T}_1$)}}
\end{picture}}
}
%\]
%--
\hspace{1.5cm}
%\[ 
\vcenter{
\hbox{
  \begin{picture}(0,0)(0,0)
\SetScale{0.4}
  \SetWidth{1.0}
\DashLine(-35,0)(-20,0){3}
\DashLine(20,0)(35,0){3}
\CArc(0,0)(20,0,180)
\CArc(0,0)(20,180,360)
\CCirc(0,20){5}{0.9}{0.9}
\CCirc(34.5,23.5){5}{0.9}{0.9}
  \SetWidth{3.5}
\CArc(60,0)(40,150,180)
\CArc(0,34.6)(40,300,330)
\CArc(30.20,17.60)(5.28,-34,153)
\Text(-16,8)[c]{\footnotesize{$t$}}
\Text(0,-20)[c]{\footnotesize{(${\mathcal T}_2$)}}
\end{picture}}
}
%\]
%--
\hspace{1.5cm}
%\[ 
\vcenter{
\hbox{
  \begin{picture}(0,0)(0,0)
\SetScale{0.4}
  \SetWidth{1.0}
\CCirc(0,20){5}{0.9}{0.9}
\CCirc(0,0){5}{0.9}{0.9}
\Line(-20,0)(20,0)
\CArc(0,0)(20,0,180)
  \SetWidth{3.5}
\DashLine(-35,0)(-20,0){3}
\DashLine(20,0)(35,0){3}
\CArc(0,0)(20,180,360)
\Text(-16,8)[c]{\footnotesize{$p_4^2$}}
\Text(0,-20)[c]{\footnotesize{(${\mathcal T}_3$)}}
\end{picture}}
}
%\]
%--
\hspace{1.5cm}
%\[ 
\vcenter{
\hbox{
  \begin{picture}(0,0)(0,0)
\SetScale{0.4}
  \SetWidth{1.0}
\CCirc(0,20){5}{0.9}{0.9}
\CCirc(0,-20){5}{0.9}{0.9}
\Line(-20,0)(20,0)
\CArc(0,0)(20,0,180)
  \SetWidth{3.5}
\DashLine(-35,0)(-20,0){3}
\DashLine(20,0)(35,0){3}
\CArc(0,0)(20,180,360)
\Text(-16,8)[c]{\footnotesize{$p_4^2$}}
\Text(0,-20)[c]{\footnotesize{(${\mathcal T}_{4}$)}}
\end{picture}}
}
%\]
%--
\hspace{1.5cm}
%\[ 
\vcenter{
\hbox{
  \begin{picture}(0,0)(0,0)
\SetScale{0.4}
  \SetWidth{1.0}
\DashLine(-35,0)(-20,0){3}
\DashLine(20,0)(35,0){3}
\CCirc(0,20){5}{0.9}{0.9}
\CCirc(0,0){5}{0.9}{0.9}
\Line(-20,0)(20,0)
\CArc(0,0)(20,0,180)
  \SetWidth{3.5}
\CArc(0,0)(20,180,360)
\Text(-16,8)[c]{\footnotesize{$t$}}
\Text(0,-20)[c]{\footnotesize{(${\mathcal T}_5$)}}
\end{picture}}
}
%\]
%--
\hspace{1.5cm}
%\[ 
\vcenter{
\hbox{
  \begin{picture}(0,0)(0,0)
\SetScale{0.4}
  \SetWidth{1.0}
\DashLine(-35,0)(-20,0){3}
\DashLine(20,0)(35,0){3}
\CCirc(0,20){5}{0.9}{0.9}
\CCirc(0,-20){5}{0.9}{0.9}
\Line(-20,0)(20,0)
\CArc(0,0)(20,0,180)
  \SetWidth{3.5}
\CArc(0,0)(20,180,360)
\Text(-16,8)[c]{\footnotesize{$t$}}
\Text(0,-20)[c]{\footnotesize{(${\mathcal T}_{6}$)}}
\end{picture}}
}
%%\[
\hspace{1.5cm}
%\[ 
\vcenter{
\hbox{
  \begin{picture}(0,0)(0,0)
\SetScale{0.4}
  \SetWidth{1.0}
\DashLine(-35,0)(-20,0){3}
\DashLine(20,0)(35,0){3}
\CArc(0,0)(20,0,180)
\CArc(0,0)(20,180,360)
\CCirc(0,20){5}{0.9}{0.9}
\CCirc(34.5,23.5){5}{0.9}{0.9}
  \SetWidth{3.5}
\CArc(60,0)(40,150,180)
\CArc(0,34.6)(40,300,330)
\CArc(30.20,17.60)(5.28,-34,153)
\Text(-16,8)[c]{\footnotesize{$s$}}
\Text(0,-20)[c]{\footnotesize{(${\mathcal T}_7$)}}
\end{picture}}
}
%\]
%--
\hspace{1.5cm}
%\[ 
\vcenter{
\hbox{
  \begin{picture}(0,0)(0,0)
\SetScale{0.4}
  \SetWidth{1.0}
%\DashLine(-35,0)(-20,0){3}
%\DashLine(20,0)(35,0){3}
\CCirc(0,20){5}{0.9}{0.9}
%\CCirc(0,-20){5}{0.9}{0.9}
\Line(-20,0)(20,0)
\CArc(0,0)(20,0,180)
  \SetWidth{3.5}
\CArc(0,0)(20,180,360)
%
%\Text(-15,10)[c]{\footnotesize{$t$}}
%
\Text(0,-20)[c]{\footnotesize{(${\mathcal T}_{8}$)}}
\end{picture}}
}
\]
\vspace*{0.3cm}
\[
\vcenter{
\hbox{
  \begin{picture}(0,0)(0,0)
\SetScale{0.4}
  \SetWidth{1.0}
\DashLine(-35,0)(-20,0){3}
\DashLine(20,0)(35,0){3}
\CCirc(0,20){5}{0.9}{0.9}
\CCirc(0,0){5}{0.9}{0.9}
\Line(-20,0)(20,0)
\CArc(0,0)(20,0,180)
  \SetWidth{3.5}
\CArc(0,0)(20,180,360)
\Text(-16,8)[c]{\footnotesize{$s$}}
\Text(0,-20)[c]{\footnotesize{(${\mathcal T}_9$)}}
\end{picture}}
}
%\]
%--
\hspace{1.5cm}
%\[ 
\vcenter{
\hbox{
  \begin{picture}(0,0)(0,0)
\SetScale{0.4}
  \SetWidth{1.0}
\DashLine(-35,0)(-20,0){3}
\DashLine(20,0)(35,0){3}
\CCirc(0,20){5}{0.9}{0.9}
\CCirc(0,-20){5}{0.9}{0.9}
\Line(-20,0)(20,0)
\CArc(0,0)(20,0,180)
  \SetWidth{3.5}
\CArc(0,0)(20,180,360)
\Text(-16,8)[c]{\footnotesize{$s$}}
\Text(0,-20)[c]{\footnotesize{(${\mathcal T}_{10}$)}}
\end{picture}}
}
%%\[
\hspace{1.5cm}
%\[
\vcenter{
\hbox{
  \begin{picture}(0,0)(0,0)
\SetScale{0.4}
  \SetWidth{1.0}
\DashLine(-45,0)(-30,0){3}
\DashLine(30,0)(45,0){3}
\CArc(-15,0)(15,0,180)
\CArc(-15,0)(15,180,360)
\CCirc(-15,15){5}{0.9}{0.9}
\CCirc(15,15){5}{0.9}{0.9}
  \SetWidth{3.5}
\CArc(15,0)(15,0,180)
\CArc(15,0)(15,180,360)
\Text(-16,8)[c]{\footnotesize{$p_4^2$}}
\Text(0,-20)[c]{\footnotesize{(${\mathcal T}_{11}$)}}
\end{picture}}
} 
%%\[
\hspace{1.5cm}
%\[
\vcenter{
\hbox{
  \begin{picture}(0,0)(0,0)
\SetScale{0.4}
  \SetWidth{1.0}
\DashLine(-45,0)(-30,0){3}
\DashLine(30,0)(45,0){3}
\CArc(-15,0)(15,0,180)
\CArc(-15,0)(15,180,360)
\CCirc(-15,15){5}{0.9}{0.9}
\CCirc(15,15){5}{0.9}{0.9}
  \SetWidth{3.5}
\CArc(15,0)(15,0,180)
\CArc(15,0)(15,180,360)
\Text(-16,8)[c]{\footnotesize{$t$}}
\Text(0,-20)[c]{\footnotesize{(${\mathcal T}_{12}$)}}
\end{picture}}
} 
%\]
%--
\hspace{1.5cm}
%\[ 
\vcenter{
\hbox{
  \begin{picture}(0,0)(0,0)
\SetScale{0.4}
  \SetWidth{1.0}
\CArc(0,0)(20,90,270)
\CArc(-20,20)(20,270,0)
%\CArc(-20,-20)(20,0,90)
%
%
  \SetWidth{3.5}
\CCirc(-15,15){4}{0.9}{0.9}
\CCirc(0,-20){4}{0.9}{0.9}
\DashLine(-35,0)(-20,0){3}
\DashLine(20,0)(35,0){3}
\CArc(0,0)(20,0,90)
\CArc(0,0)(20,180,360)
%\Line(0,-20)(0,20)
\Text(-16,8)[c]{\footnotesize{$p_4^2$}}
\Text(0,-20)[c]{\footnotesize{(${\mathcal T}_{13}$)}}
\end{picture}}
}
%%\[
\hspace{1.5cm}
%\[
\vcenter{
\hbox{
  \begin{picture}(0,0)(0,0)
\SetScale{0.4}
  \SetWidth{1.0}
\DashLine(-45,0)(-30,0){3}
\DashLine(30,0)(45,0){3}
\CArc(-15,0)(15,0,180)
\CArc(-15,0)(15,180,360)
\CCirc(-15,15){5}{0.9}{0.9}
\CCirc(15,15){5}{0.9}{0.9}
  \SetWidth{3.5}
\CArc(15,0)(15,0,180)
\CArc(15,0)(15,180,360)
\Text(-16,8)[c]{\footnotesize{$s$}}
\Text(0,-20)[c]{\footnotesize{(${\mathcal T}_{14}$)}}
\end{picture}}
} 
%\]
%--
\hspace{1.5cm}
%\[ 
%%\]
%%
%%\vspace*{2.0cm}
%%
%%\[
\vcenter{
\hbox{
  \begin{picture}(0,0)(0,0)
\SetScale{0.4}
  \SetWidth{1}
\DashLine(-50,0)(-35,0){3}
\Line(10,30)(25,30)
\CArc(-25,34)(35,255,350)
\Line(-35,0)(10,30)
  \SetWidth{3.5}
\CCirc(-13,15){5}{0.9}{0.9}
\DashLine(10,-30)(25,-30){3}
\Line(10,30)(10,-30)
\Line(10,-30)(-35,0)
\Text(-16,8)[c]{\footnotesize{$s$}}
\Text(0,-25)[c]{\footnotesize{(${\mathcal T}_{15}$)}}
\end{picture}}
}
%\]
%--
\hspace{1.5cm}
%\[ 
%%\[
\vcenter{
\hbox{
  \begin{picture}(0,0)(0,0)
\SetScale{0.4}
  \SetWidth{1}
\DashLine(-50,0)(-35,0){3}
\Line(10,30)(25,30)
\CArc(-25,34)(35,255,350)
\Line(-35,0)(10,30)
  \SetWidth{3.5}
\CCirc(-13,15){5}{0.9}{0.9}
\CCirc(-13,-15){5}{0.9}{0.9}
\DashLine(10,-30)(25,-30){3}
\Line(10,30)(10,-30)
\Line(10,-30)(-35,0)
\Text(-16,8)[c]{\footnotesize{$s$}}
%\Text(28,-15)[c]{$m_H^2$}
% \SetWidth{.4}
\Text(0,-25)[c]{\footnotesize{(${\mathcal T}_{16}$)}}
\end{picture}}
}
\]
\vspace*{0.5cm}
\[
\vcenter{
\hbox{
  \begin{picture}(0,0)(0,0)
\SetScale{0.4}
  \SetWidth{1}
\DashLine(-50,0)(-35,0){3}
\Line(10,30)(25,30)
\CArc(-25,34)(35,255,350)
\Line(-35,0)(10,30)
%\CCirc(-13,15){5}{0.9}{0.9}
%
  \SetWidth{3.5}
\CCirc(-13,15){5}{0.9}{0.9}
\DashLine(10,-30)(25,-30){3}
\Line(10,30)(10,-30)
\Line(10,-30)(-35,0)
\Text(-16,8)[c]{\footnotesize{$t$}}
%\Text(28,-15)[c]{$m_H^2$}
% \SetWidth{.4}
\Text(0,-28)[c]{\footnotesize{(${\mathcal T}_{17}$)}}
\end{picture}}
}
%\]
%--
\hspace{1.5cm}
%\[ 
%%\[
\vcenter{
\hbox{
  \begin{picture}(0,0)(0,0)
\SetScale{0.4}
  \SetWidth{1}
\DashLine(-50,0)(-35,0){3}
\Line(10,30)(25,30)
\CArc(-25,34)(35,255,350)
\Line(-35,0)(10,30)
%\CCirc(-13,15){5}{0.9}{0.9}
%
  \SetWidth{3.5}
\CCirc(-13,15){5}{0.9}{0.9}
\CCirc(-13,-15){5}{0.9}{0.9}
\DashLine(10,-30)(25,-30){3}
\Line(10,30)(10,-30)
\Line(10,-30)(-35,0)
\Text(-16,8)[c]{\footnotesize{$t$}}
%\Text(28,-15)[c]{$m_H^2$}
% \SetWidth{.4}
\Text(0,-28)[c]{\footnotesize{(${\mathcal T}_{18}$)}}
\end{picture}}
}
%\]
%--
\hspace{1.5cm}
%\[ 
%%\]
\vcenter{
\hbox{
  \begin{picture}(0,0)(0,0)
\SetScale{0.4}
  \SetWidth{1}
\DashLine(-50,0)(-35,0){3}
\Line(10,30)(25,30)
\CArc(30,0)(35,125,235)
\CCirc(-5,0){5}{0.9}{0.9}
\Line(-35,0)(10,30)
\Line(10,-30)(-35,0)
  \SetWidth{3.5}
\DashLine(10,-30)(25,-30){3}
\Line(10,30)(10,-30)
\Text(-16,8)[c]{\footnotesize{$t$}}
%\Text(28,-15)[c]{$m_H^2$}
\Text(0,-28)[c]{\footnotesize{(${\mathcal T}_{19}$)}}
\end{picture}}
}
%\]
%--
\hspace{1.5cm}
%\[ 
\vcenter{
\hbox{
  \begin{picture}(0,0)(0,0)
\SetScale{0.4}
  \SetWidth{1}
\DashLine(-50,0)(-35,0){3}
\Line(10,30)(25,30)
\CArc(30,0)(35,125,235)
\CCirc(10,0){5}{0.9}{0.9}
\CCirc(-5,0){5}{0.9}{0.9}
\Line(-35,0)(10,30)
\Line(10,-30)(-35,0)
  \SetWidth{3.5}
\DashLine(10,-30)(25,-30){3}
\Line(10,30)(10,-30)
\Text(-16,8)[c]{\footnotesize{$t$}}
%\Text(28,-15)[c]{$m_H^2$}
\Text(0,-28)[c]{\footnotesize{(${\mathcal T}_{20}$)}}
\end{picture}}
}
%\]
%--
\hspace{1.5cm}
%\[ 
%%\]
\vcenter{
\hbox{
  \begin{picture}(0,0)(0,0)
\SetScale{0.4}
  \SetWidth{1}
\DashLine(-50,0)(-35,0){3}
\Line(10,30)(25,30)
\CArc(30,0)(35,125,235)
\CCirc(-5,0){5}{0.9}{0.9}
\Line(-35,0)(10,30)
\Line(10,-30)(-35,0)
  \SetWidth{3.5}
\DashLine(10,-30)(25,-30){3}
\Line(10,30)(10,-30)
\Text(-16,8)[c]{\footnotesize{$s$}}
%\Text(28,-15)[c]{$m_H^2$}
\Text(0,-28)[c]{\footnotesize{(${\mathcal T}_{21}$)}}
\end{picture}}
}
%\]
%--
\hspace{1.5cm}
%\[ 
\vcenter{
\hbox{
  \begin{picture}(0,0)(0,0)
\SetScale{0.4}
  \SetWidth{1}
\DashLine(-50,0)(-35,0){3}
\Line(10,30)(25,30)
\CArc(30,0)(35,125,235)
\CCirc(10,0){5}{0.9}{0.9}
\CCirc(-5,0){5}{0.9}{0.9}
\Line(-35,0)(10,30)
\Line(10,-30)(-35,0)
  \SetWidth{3.5}
\DashLine(10,-30)(25,-30){3}
\Line(10,30)(10,-30)
\Text(-16,8)[c]{\footnotesize{$s$}}
%\Text(28,-15)[c]{$m_H^2$}
\Text(0,-28)[c]{\footnotesize{(${\mathcal T}_{22}$)}}
\end{picture}}
}
%\]
%--
\hspace{1.5cm}
%\[ 
\vcenter{
\hbox{
  \begin{picture}(0,0)(0,0)
\SetScale{0.4}
  \SetWidth{1.0}
\CArc(0,0)(20,90,270)
%\CArc(-20,-20)(20,0,90)
%\CCirc(-6,-6){4}{0.9}{0.9}
%
%
\Line(0,-20)(0,20)
  \SetWidth{3.5}
\CCirc(13,14){4}{0.9}{0.9}
\DashLine(-35,0)(-20,0){3}
\DashLine(20,0)(35,0){3}
\CArc(0,0)(20,0,90)
\CArc(0,0)(20,270,360)
%\Line(0,-20)(0,20)
\Text(-16,8)[c]{\footnotesize{$p_4^2$}}
\Text(0,-28)[c]{\footnotesize{(${\mathcal T}_{23}$)}}
\end{picture}}
}
%\]
%--
\hspace{1.5cm}
%\[ 
\vcenter{
\hbox{
  \begin{picture}(0,0)(0,0)
\SetScale{0.4}
  \SetWidth{1}
\DashLine(-50,0)(-35,0){3}
\Line(10,30)(25,30)
%\CArc(30,0)(35,125,235)
%
%\CCirc(10,0){5}{0.9}{0.9}
%\CCirc(-13,15){5}{0.9}{0.9}
%
\Line(10,30)(10,-30)
\Line(-35,0)(10,30)
\Line(-35,0)(10,0)
  \SetWidth{3.5}
\DashLine(10,-30)(25,-30){3}
\Line(10,0)(10,-30)
\Line(10,-30)(-35,0)
\Text(-16,8)[c]{\footnotesize{$t$}}
%\Text(28,-15)[c]{$m_H^2$}
\Text(0,-28)[c]{\footnotesize{(${\mathcal T}_{24}$)}}
\end{picture}}
}
%\]
%--
%\]
\]
\vspace*{0.7cm}
\[
\vcenter{
\hbox{
  \begin{picture}(0,0)(0,0)
\SetScale{0.4}
  \SetWidth{1}
\Line(-25,30)(-40,30)
\Line(25,30)(40,30)
\Line(25,30)(-25,30)
\Line(-25,-30)(-40,-30)
\Line(-25,30)(-25,-30)
\Line(-25,-30)(25,-30)
\Line(25,-30)(25,30)
%\Line(0,0)(0,30)
%
\CCirc(40,-10){5}{0.9}{0.9}
  \SetWidth{3.5}
\DashLine(25,-30)(25,-45){3}
\CArc(40,-25)(15,0,180)
\CArc(40,-25)(15,180,360)
%\CArc(42,-40)(20,30,150)
%\CArc(42,-20)(20,210,330)
%
%
%
\Text(0,-28)[c]{\footnotesize{(${\mathcal T}_{25}$)}}
\end{picture}}
}
%--
\hspace{1.5cm}
%\[ 
\vcenter{
\hbox{
  \begin{picture}(0,0)(0,0)
\SetScale{0.4}
  \SetWidth{1}
\DashLine(-50,0)(-35,0){3}
\Line(10,30)(25,30)
%\CArc(30,0)(35,125,235)
%
%\CCirc(10,0){5}{0.9}{0.9}
%\CCirc(-13,15){5}{0.9}{0.9}
%
\Line(10,30)(10,-30)
\Line(-35,0)(10,30)
\Line(-35,0)(10,0)
  \SetWidth{3.5}
\DashLine(10,-30)(25,-30){3}
\Line(10,0)(10,-30)
\Line(10,-30)(-35,0)
\Text(-16,8)[c]{\footnotesize{$s$}}
%\Text(28,-15)[c]{$m_H^2$}
\Text(0,-28)[c]{\footnotesize{(${\mathcal T}_{26}$)}}
\end{picture}}
}
%--
\hspace{1.5cm}
%\[ 
\vcenter{
\hbox{
  \begin{picture}(0,0)(0,0)
\SetScale{0.4}
  \SetWidth{1.0}
\DashLine(-50,0)(-35,0){3}
\Line(10,30)(25,30)
\Line(-35,0)(10,30)
\Line(-11,-16)(-35,0)
%\CCirc(-12,15){5}{0.9}{0.9}
%\CCirc(10,0){5}{0.9}{0.9}
%
\Line(10,30)(-11,-16)
  \SetWidth{3.5}
\DashLine(10,-30)(25,-30){3}
%\Line(11,16)(35,0)
\Line(10,-30)(-11,-16)
\Line(10,30)(10,-30)
\Text(-16,8)[c]{\footnotesize{$t$}}
\Text(0,-28)[c]{\footnotesize{(${\mathcal T}_{27}$)}}
\end{picture}}
}
%--
\hspace{1.5cm}
%\[ 
\vcenter{
\hbox{
  \begin{picture}(0,0)(0,0)
\SetScale{0.4}
  \SetWidth{1.0}
\DashLine(-50,0)(-35,0){3}
\Line(10,30)(25,30)
\Line(-35,0)(10,30)
\Line(-11,-16)(-35,0)
%\CCirc(-12,15){5}{0.9}{0.9}
\CCirc(-3,0){5}{0.9}{0.9}
\Line(10,30)(-11,-16)
  \SetWidth{3.5}
\DashLine(10,-30)(25,-30){3}
%\Line(11,16)(35,0)
\Line(10,-30)(-11,-16)
\Line(10,30)(10,-30)
\Text(-16,8)[c]{\footnotesize{$t$}}
\Text(0,-28)[c]{\footnotesize{(${\mathcal T}_{28}$)}}
\end{picture}}
}
%--
\hspace{1.5cm}
%\[ 
\vcenter{
\hbox{
  \begin{picture}(0,0)(0,0)
\SetScale{0.4}
  \SetWidth{1.0}
\DashLine(-50,0)(-35,0){3}
\Line(10,30)(25,30)
\Line(-35,0)(10,30)
\Line(-11,-16)(-35,0)
\CCirc(-20,-8){5}{0.9}{0.9}
\CCirc(10,0){5}{0.9}{0.9}
\Line(10,30)(-11,-16)
  \SetWidth{3.5}
\DashLine(10,-30)(25,-30){3}
%\Line(11,16)(35,0)
\Line(10,-30)(-11,-16)
\Line(10,30)(10,-30)
\Text(-16,8)[c]{\footnotesize{$t$}}
\Text(0,-28)[c]{\footnotesize{(${\mathcal T}_{29}$)}}
\end{picture}}
}
%--
\hspace{1.5cm}
%\[ 
\vcenter{
\hbox{
  \begin{picture}(0,0)(0,0)
\SetScale{0.4}
  \SetWidth{1.0}
\DashLine(-50,0)(-35,0){3}
\Line(10,30)(25,30)
\Line(-35,0)(10,30)
\Line(-11,-16)(-35,0)
%\CCirc(-12,15){5}{0.9}{0.9}
\CCirc(10,0){5}{0.9}{0.9}
\Line(10,30)(-11,-16)
  \SetWidth{3.5}
\DashLine(10,-30)(25,-30){3}
%\Line(11,16)(35,0)
\Line(10,-30)(-11,-16)
\Line(10,30)(10,-30)
\Text(-16,8)[c]{\footnotesize{$t$}}
\Text(0,-28)[c]{\footnotesize{(${\mathcal T}_{30}$)}}
\end{picture}}
}
%--
\hspace{1.5cm}
%\[ 
\vcenter{
\hbox{
  \begin{picture}(0,0)(0,0)
\SetScale{0.4}
  \SetWidth{1.0}
\DashLine(-50,0)(-35,0){3}
\Line(10,30)(25,30)
\Line(-35,0)(10,30)
\Line(-11,-16)(-35,0)
%\CCirc(-12,15){5}{0.9}{0.9}
%\CCirc(10,0){5}{0.9}{0.9}
%
\Line(10,30)(-11,-16)
  \SetWidth{3.5}
\DashLine(10,-30)(25,-30){3}
%\Line(11,16)(35,0)
\Line(10,-30)(-11,-16)
\Line(10,30)(10,-30)
\Text(-16,8)[c]{\footnotesize{$s$}}
\Text(0,-28)[c]{\footnotesize{(${\mathcal T}_{31}$)}}
\end{picture}}
}
%--
\hspace{1.5cm}
%\[ 
\vcenter{
\hbox{
  \begin{picture}(0,0)(0,0)
\SetScale{0.4}
  \SetWidth{1.0}
\DashLine(-50,0)(-35,0){3}
\Line(10,30)(25,30)
\Line(-35,0)(10,30)
\Line(-11,-16)(-35,0)
%\CCirc(-12,15){5}{0.9}{0.9}
\CCirc(-15,13){5}{0.9}{0.9}
\Line(10,30)(-11,-16)
  \SetWidth{3.5}
\DashLine(10,-30)(25,-30){3}
%\Line(11,16)(35,0)
\Line(10,-30)(-11,-16)
\Line(10,30)(10,-30)
\Text(-16,8)[c]{\footnotesize{$s$}}
\Text(0,-28)[c]{\footnotesize{(${\mathcal T}_{32}$)}}
\end{picture}}
}
\]
\vspace*{0.7cm}
\[
\vcenter{
\hbox{
  \begin{picture}(0,0)(0,0)
\SetScale{0.4}
  \SetWidth{1.0}
\DashLine(-50,0)(-35,0){3}
\Line(10,30)(25,30)
\Line(-35,0)(10,30)
\Line(-11,-16)(-35,0)
\CCirc(-20,-8){5}{0.9}{0.9}
\CCirc(10,0){5}{0.9}{0.9}
\Line(10,30)(-11,-16)
  \SetWidth{3.5}
\DashLine(10,-30)(25,-30){3}
%\Line(11,16)(35,0)
\Line(10,-30)(-11,-16)
\Line(10,30)(10,-30)
\Text(-16,8)[c]{\footnotesize{$s$}}
\Text(0,-28)[c]{\footnotesize{(${\mathcal T}_{33}$)}}
\end{picture}}
}
%--
\hspace{1.5cm}
%\[ 
\vcenter{
\hbox{
  \begin{picture}(0,0)(0,0)
\SetScale{0.4}
  \SetWidth{1.0}
\DashLine(-50,0)(-35,0){3}
\Line(10,30)(25,30)
\Line(-35,0)(10,30)
\Line(-11,-16)(-35,0)
%\CCirc(-12,15){5}{0.9}{0.9}
\CCirc(10,0){5}{0.9}{0.9}
\Line(10,30)(-11,-16)
  \SetWidth{3.5}
\DashLine(10,-30)(25,-30){3}
%\Line(11,16)(35,0)
\Line(10,-30)(-11,-16)
\Line(10,30)(10,-30)
\Text(-16,8)[c]{\footnotesize{$s$}}
\Text(0,-28)[c]{\footnotesize{(${\mathcal T}_{34}$)}}
\end{picture}}
}
%--
\hspace{1.5cm}
%\[ 
%\[ 
\vcenter{
\hbox{
  \begin{picture}(0,0)(0,0)
\SetScale{0.4}
  \SetWidth{1}
\Line(-25,30)(-40,30)
\Line(-25,-30)(25,30)
\Line(25,30)(40,30)
\Line(-25,-30)(-40,-30)
%\Line(0,0)(0,30)
%
%\CCirc(25,0){5}{0.9}{0.9}
%
\Line(-25,30)(-25,-30)
\Line(25,30)(-25,30)
  \SetWidth{3.5}
\DashLine(25,-30)(40,-30){3}
\Line(-25,-30)(25,-30)
\Line(25,-30)(25,30)
\Text(0,-28)[c]{\footnotesize{(${\mathcal T}_{35}$)}}
\end{picture}}
}
%--
\hspace{1.5cm}
%\[ 
\vcenter{
\hbox{
  \begin{picture}(0,0)(0,0)
\SetScale{0.4}
  \SetWidth{1}
\Line(-25,30)(-40,30)
\Line(-25,-30)(25,30)
\Line(25,30)(40,30)
\Line(-25,-30)(-40,-30)
%\Line(0,0)(0,30)
%
\CCirc(0,0){5}{0.9}{0.9}
\Line(-25,30)(-25,-30)
\Line(25,30)(-25,30)
  \SetWidth{3.5}
\DashLine(25,-30)(40,-30){3}
\Line(-25,-30)(25,-30)
\Line(25,-30)(25,30)
\Text(0,-28)[c]{\footnotesize{(${\mathcal T}_{36}$)}}
\end{picture}}
}
%--
\hspace{1.5cm}
%\[
\vcenter{
\hbox{
  \begin{picture}(0,0)(0,0)
\SetScale{0.4}
  \SetWidth{1}
\Line(-25,30)(-40,30)
%\Line(-25,-30)(25,30)
\Line(25,30)(40,30)
\Line(-25,-30)(-40,-30)
%\Line(0,0)(0,30)
%
\CCirc(0,-30){5}{0.9}{0.9}
\Line(-25,30)(-25,-30)
\Line(-25,-30)(25,-30)
\Line(25,-30)(25,30)
\Line(25,30)(-25,30)
  \SetWidth{3.5}
\DashLine(25,-30)(40,-30){3}
\CArc(0,-50)(33,40,140)
\Text(0,-28)[c]{\footnotesize{(${\mathcal T}_{37}$)}}
\end{picture}}
}
%--
\hspace{1.5cm}
%\[ 
\vcenter{
\hbox{
  \begin{picture}(0,0)(0,0)
\SetScale{0.4}
  \SetWidth{1}
\Line(-25,30)(-40,30)
%\Line(-25,-30)(25,30)
\Line(25,30)(40,30)
\Line(-25,-30)(-40,-30)
%\Line(0,0)(0,30)
%
\CCirc(0,-17){5}{0.9}{0.9}
\Line(-25,30)(-25,-30)
\Line(-25,-30)(25,-30)
\Line(25,-30)(25,30)
\Line(25,30)(-25,30)
  \SetWidth{3.5}
\DashLine(25,-30)(40,-30){3}
\CArc(0,-50)(33,40,140)
\Text(0,-28)[c]{\footnotesize{(${\mathcal T}_{38}$)}}
\end{picture}}
}
%--
\hspace{1.5cm}
%\[ 
\vcenter{
\hbox{
  \begin{picture}(0,0)(0,0)
\SetScale{0.4}
  \SetWidth{1}
\Line(-25,30)(-40,30)
\Line(25,30)(40,30)
\Line(-25,-30)(-40,-30)
\Line(-25,30)(-25,-30)
\Line(-25,-30)(25,-30)
\Line(25,30)(-25,30)
%\Line(0,0)(0,30)
%
\CCirc(25,0){5}{0.9}{0.9}
\Line(25,-30)(25,30)
  \SetWidth{3.5}
\DashLine(25,-30)(40,-30){3}
\CArc(45,0)(35,125,235)
\Text(0,-28)[c]{\footnotesize{(${\mathcal T}_{39}$)}}
\end{picture}}
} 
%--
\hspace{1.5cm}
%\[
\vcenter{
\hbox{
  \begin{picture}(0,0)(0,0)
\SetScale{0.4}
  \SetWidth{1}
\Line(-25,30)(-40,30)
\Line(25,30)(40,30)
\Line(-25,-30)(-40,-30)
\Line(-25,30)(-25,-30)
\Line(-25,-30)(25,-30)
\Line(25,30)(-25,30)
%\Line(0,0)(0,30)
%
\CCirc(10,0){5}{0.9}{0.9}
\Line(25,-30)(25,30)
  \SetWidth{3.5}
\DashLine(25,-30)(40,-30){3}
\CArc(45,0)(35,125,235)
\Text(0,-28)[c]{\footnotesize{(${\mathcal T}_{40}$)}}
\end{picture}}
} 
\]
\vspace*{0.8cm}
\[
\vcenter{
\hbox{
  \begin{picture}(0,0)(0,0)
\SetScale{0.4}
  \SetWidth{1}
\Line(-25,30)(-40,30)
\Line(25,30)(40,30)
\Line(-25,-30)(-40,-30)
\Line(-25,30)(-25,-30)
\Line(-25,-30)(25,-30)
\Line(25,30)(-25,30)
%\Line(0,0)(0,30)
%
%\CCirc(10,0){5}{0.9}{0.9}
%
\Line(25,-30)(25,30)
  \SetWidth{3.5}
\DashLine(25,-30)(40,-30){3}
\CArc(10,-30)(15,0,180)
\Line(-5,-30)(25,-30)
\Text(0,-28)[c]{\footnotesize{(${\mathcal T}_{41}$)}}
\end{picture}}
}
%--
\hspace{1.5cm}
%\[ 
\vcenter{
\hbox{
  \begin{picture}(0,0)(0,0)
\SetScale{0.4}
  \SetWidth{1}
\Line(-25,30)(-40,30)
\Line(25,30)(40,30)
\Line(-25,-30)(-40,-30)
%
%\CCirc(25,0){5}{0.9}{0.9}
%
\Line(25,-30)(25,30)
\Line(-25,30)(-25,-30)
\Line(25,30)(-25,30)
\Line(-25,-30)(25,0)
  \SetWidth{3.5}
\DashLine(25,-30)(40,-30){3}
\Line(-25,-30)(25,-30)
\Line(25,-30)(25,0)
\Text(0,-28)[c]{\footnotesize{(${\mathcal T}_{42}$)}}
%
%\Text(55,-15)[c]{$m_H^2$}
\end{picture}}
}
%--
\hspace{1.5cm}
%\[
\vcenter{
\hbox{
  \begin{picture}(0,0)(0,0)
\SetScale{0.4}
  \SetWidth{1}
\Line(-25,30)(-40,30)
\Line(25,30)(40,30)
\Line(-25,-30)(-40,-30)
%
%\CCirc(25,0){5}{0.9}{0.9}
%
\Line(25,-30)(25,30)
\Line(-25,30)(-25,-30)
\Line(25,30)(-25,30)
\Line(-25,-30)(25,0)
  \SetWidth{3.5}
\DashLine(25,-30)(40,-30){3}
\Line(-25,-30)(25,-30)
\Line(25,-30)(25,0)
\Text(0,-28)[c]{\footnotesize{(${\mathcal T}_{43}$)}}
\Text(0,20)[c]{\tiny{$-(k_1\!\!+\!\!p_1\!\!+p_2)^2$}}
%
%\Text(55,-15)[c]{$m_H^2$}
\end{picture}}
} 
%--
\hspace{1.5cm}
%\[
\vcenter{
\hbox{
  \begin{picture}(0,0)(0,0)
\SetScale{0.4}
  \SetWidth{1}
\Line(-25,30)(-40,30)
\Line(25,30)(40,30)
\Line(-25,-30)(-40,-30)
\Line(25,-30)(25,30)
\Line(-25,30)(-25,-30)
\Line(25,30)(-25,30)
\Line(-25,-30)(25,0)
  \SetWidth{3.5}
\CCirc(5,-28){5}{0.9}{0.9}
\DashLine(25,-30)(40,-30){3}
\Line(-25,-30)(25,-30)
\Line(25,-30)(25,0)
\Text(0,-28)[c]{\footnotesize{(${\mathcal T}_{44}$)}}
\end{picture}}
}
%--
\hspace{1.5cm}
%\[
\vcenter{
\hbox{
  \begin{picture}(0,0)(0,0)
\SetScale{0.4}
  \SetWidth{1}
\Line(-25,30)(-40,30)
\Line(25,30)(40,30)
\Line(-25,-30)(-40,-30)
\Line(25,-30)(25,30)
\Line(-25,30)(-25,-30)
\Line(25,30)(-25,30)
\Line(25,-30)(-25,-30)
\Line(0,-30)(25,30)
  \SetWidth{3.5}
%\CCirc(5,-28){5}{0.9}{0.9}
%
\DashLine(25,-30)(40,-30){3}
\Line(0,-30)(25,-30)
\Line(25,-30)(25,30)
\Text(0,-28)[c]{\footnotesize{(${\mathcal T}_{45}$)}}
%
%\Text(55,-15)[c]{$m_H^2$}
\end{picture}}
}
%--
\hspace{1.5cm}
%\[
\vcenter{
\hbox{
  \begin{picture}(0,0)(0,0)
\SetScale{0.4}
  \SetWidth{1}
\Line(-25,30)(-40,30)
\Line(25,30)(40,30)
\Line(-25,-30)(-40,-30)
\Line(25,-30)(25,30)
\Line(-25,30)(-25,-30)
\Line(25,30)(-25,30)
\Line(25,-30)(-25,-30)
\Line(0,-30)(25,30)
  \SetWidth{3.5}
%\CCirc(5,-28){5}{0.9}{0.9}
%
\DashLine(25,-30)(40,-30){3}
\Line(0,-30)(25,-30)
\Line(25,-30)(25,30)
\Text(0,-28)[c]{\footnotesize{(${\mathcal T}_{46}$)}}
\Text(0,20)[c]{\tiny{$-(k_1\!\!-\!\!k_2\!\!+\!\!p_3)^2$}}
%
%\Text(55,-15)[c]{$m_H^2$}
\end{picture}}
}
%--
\hspace{1.5cm}
%\[
\vcenter{
\hbox{
  \begin{picture}(0,0)(0,0)
\SetScale{0.4}
  \SetWidth{1}
\Line(-25,30)(-40,30)
\Line(25,30)(40,30)
\Line(-25,-30)(-40,-30)
\Line(25,-30)(25,30)
\Line(-25,30)(-25,-30)
\Line(25,30)(-25,30)
\Line(25,-30)(-25,-30)
\Line(0,-30)(25,30)
  \SetWidth{3.5}
%\CCirc(5,-28){5}{0.9}{0.9}
%
\DashLine(25,-30)(40,-30){3}
\Line(0,-30)(25,-30)
\Line(25,-30)(25,30)
\Text(0,-28)[c]{\footnotesize{(${\mathcal T}_{47}$)}}
\end{picture}}
}
%--
\hspace{1.5cm}
%\[
\vcenter{
\hbox{
  \begin{picture}(0,0)(0,0)
\SetScale{0.4}
  \SetWidth{1}
\Line(-25,30)(-40,30)
\Line(25,30)(40,30)
\Line(-25,-30)(-40,-30)
%\Line(0,-30)(0,30)
\Line(-25,30)(-25,-30)
\Line(-25,-30)(25,-30)
\Line(25,-30)(25,30)
\Line(25,30)(-25,30)
\Line(25,0)(0,-30)
%\CCirc(40,-10){5}{0.9}{0.9}
%
  \SetWidth{3.5}
\DashLine(25,-30)(40,-30){3}
\Line(25,-30)(25,0)
\Line(25,-30)(0,-30)
\Text(0,-28)[c]{\footnotesize{(${\mathcal T}_{48}$)}}
\end{picture}}
}
\]
\vspace*{7mm}
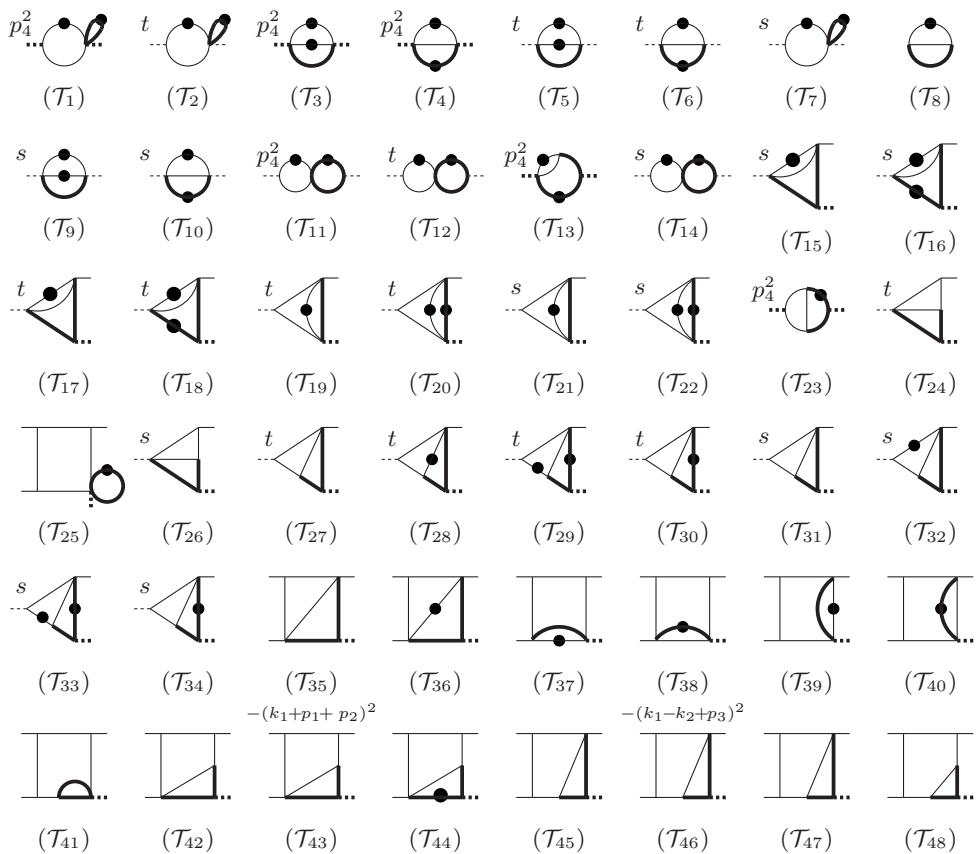
\caption{Master integrals in pre-canonical form. Internal plain thin lines represent massless
propagators, while thick lines represent the top propagator. External plain thin lines
represent massless particles on their mass-shell. External dashed thin lines represent the
dependence on $s$, $t$, or $m_H^2$. The external dashed thick line represents the Higgs on its
mass-shell. \label{figmasterprecan}}
\ec
\efig
%%%%%%%%%%%%%%%%%%%%%%%%%%%%%%%%%%%%%%%%%%%%%%%%%%%%%%%%%%%%%%%%%%%%%%%%
%

%%%%%%%%%%%%%%%%%%%%%%%%%%%%%%%%%%%%%%

\section{Conclusions}
\label{Sec5}

In this paper we have computed analytically the planar master integrals relevant to the evaluation of the 
two-loop four-point amplitudes for Higgs plus three gluons, where one loop is an electroweak-boson loop and the other is a light-quark loop.
Those amplitudes contribute to the mixed QCD-electroweak light-fermion corrections to the production of a Higgs boson with an additional jet,
as well as to the real radiation of the NLO QCD corrections to the two-loop light-quark contribution to Higgs production via gluon-gluon fusion.

The master integrals are evaluated with the differential equations method applied to a canonical set of basis integrals. Since the alphabet of the solution depends on a single square root, it is possible to find a variable change such that the matrices associated to the system of 
differential equations can be expressed in terms of rational functions. This allows a direct integration of the differential equations in terms of generalized polylogarithms 
up to weight 4. The expression of the master integrals in terms of generalized polylogarithms is quite flexible and can be
evaluated numerically in a fast and precise way.

%%%%%%%%%%%%%%%%%%%%%%%%%%%%%%%%%%%%%%

\section{Acknowledgments}

MB and VC thank the Institut f\"{u}r Theoretische Physik of the ETH Z\"{u}rich for the hospitality and the COST
(European Cooperation in Science and Technology) Action CA16201 PARTICLEFACE for the support, during the early stages of this work.

\appendix

\appendix

\section{Routing for the Pre-Canonical Master Integrals}

In this appendix we give the expression for the pre-canonical master integrals in the form 
of Eq.~(\ref{integral}).

\begin{align}
\T_{1}&=\intd\frac{1}{D_{4}^2 D_{5} D_{7}^{2}},\Space \T_{2}=\intd\frac{1}{D_{3}^{2}D_{5}D_{7}^{2}}, \nonumber\\
\T_{3}&=\intd\frac{1}{D_{2}^{2}D_{5}^{2}D_{6}},\Space 
\T_{4}=\intd\frac{1}{D_{2}^{2}D_{5}D_{6}^{2}}, \nonumber\\
\T_{5}&=\intd\frac{1}{D_{2}^{2}D_{3}^{2}D_{7}},\Space 
\T_{6}=\intd\frac{1}{D_{2}^{2}D_{3}D_{7}^{2}}, \nonumber\\
\T_{7}&=\intd\frac{1}{D_{1}^{2}D_{4}D_{7}^{2}},\Space
\T_{8}=\intd\frac{1}{D_{1}D_{2}D_{7}}, \nonumber\\
\T_{9}&=\intd\frac{1}{D_{1}^{2}D_{2}^{2}D_{6}},\Space
\T_{10}=\intd\frac{1}{D_{1}^{2}D_{2}D_{6}^{2}}, \nonumber\\
\T_{11}&=\intd\frac{1}{D_{4}^{2}D_{5}D_{6}D_{7}^{2}},\qquad\qquad\quad
\T_{12}=\intd\frac{1}{D_{3}^{2}D_{5}D_{6}^{2}D_{7}}, \nonumber\\
\T_{13}&=\intd\frac{1}{D_{2}^{2}D_{5}D_{6}^{2}D_{7}},\qquad\qquad\quad
\T_{14}=\intd\frac{1}{D_{1}^{2}D_{4}D_{6}D_{7}^{2}}, \nonumber\\
\T_{15}&=\intd\frac{1}{D_{1}^{2}D_{2}D_{6}D_{7}},\qquad\qquad\quad
\T_{16}=\intd\frac{1}{D_{1}^{2}D_{2}D_{6}^{2}D_{7}}, \nonumber\\
\T_{17}&=\intd\frac{1}{D_{2}^{2}D_{3}D_{6}D_{7}},\qquad\qquad\quad
\T_{18}=\intd\frac{1}{D_{2}^{2}D_{3}D_{6}D_{7}^{2}}, \nonumber\\
\T_{19}&=\intd\frac{1}{D_{2}^{2}D_{3}D_{5}D_{6}},\qquad\qquad\quad
\T_{20}=\intd\frac{1}{D_{2}^{2}D_{3}D_{5}D_{6}^{2}}, \nonumber\\
\T_{21}&=\intd\frac{1}{D_{1}D_{2}^{2}D_{4}D_{7}},\qquad\qquad\quad
\T_{22}=\intd\frac{1}{D_{1}D_{2}^{2}D_{4}D_{7}^{2}}, \nonumber\\
\T_{23}&=\intd\frac{1}{D_{2}D_{4}D_{5}D_{6}D_{7}^{2}},\qquad\quad\quad
\T_{24}=\intd\frac{1}{D_{2}D_{3}D_{4}D_{6}D_{7}}, \nonumber\\
\T_{25}&=\intd\frac{1}{D_{1}D_{3}D_{4}D_{5}D_{7}^{2}},\qquad\quad\quad
\T_{26}=\intd\frac{1}{D_{1}D_{2}D_{5}D_{6}D_{7}}, \nonumber\\
\T_{27}&=\intd\frac{1}{D_{2}D_{3}D_{5}D_{6}D_{7}},\qquad\quad\quad
\T_{28}=\intd\frac{1}{D_{2}^{2}D_{3}D_{5}D_{6}D_{7}}, \nonumber\\
\T_{29}&=\intd\frac{1}{D_{2}D_{3}D_{5}^{2}D_{6}^{2}D_{7}},\qquad\quad\quad
\T_{30}=\intd\frac{1}{D_{2}D_{3}D_{5}D_{6}^{2}D_{7}}, \nonumber\\
\T_{31}&=\intd\frac{1}{D_{1}D_{2}D_{4}D_{6}D_{7}},\qquad\quad\quad
\T_{32}=\intd\frac{1}{D_{1}^{2}D_{2}D_{4}D_{6}D_{7}}, \nonumber\\
\T_{33}&=\intd\frac{1}{D_{1}D_{2}D_{4}^{2}D_{6}D_{7}^{2}},\qquad\quad\quad
\T_{34}=\intd\frac{1}{D_{1}D_{2}D_{4}D_{6}D_{7}^{2}}, \nonumber\\
\T_{35}&=\intd\frac{1}{D_{1}D_{2}D_{3}D_{6}D_{7}},\qquad\quad\quad
\T_{36}=\intd\frac{1}{D_{1}D_{2}^{2}D_{3}D_{6}D_{7}}, \nonumber\\
\T_{37}&=\intd\frac{1}{D_{1}D_{2}^{2}D_{3}D_{5}D_{6}},\qquad\quad\quad
\T_{38}=\intd\frac{1}{D_{1}D_{2}D_{3}D_{5}D_{6}^{2}}, \nonumber\\
\T_{39}&=\intd\frac{1}{D_{1}D_{2}D_{3}D_{4}D_{7}},\qquad\quad\quad
\T_{40}=\intd\frac{1}{D_{1}D_{2}D_{3}D_{4}D_{7}^{2}}, \nonumber\\
\T_{41}&=\intd\frac{1}{D_{1}D_{3}D_{4}D_{5}D_{6}D_{7}^{2}},\quad\quad\;\;
\T_{42}=\intd\frac{1}{D_{1}D_{2}D_{3}D_{5}D_{6}D_{7}}, \nonumber\\
\T_{43}&=\intd\frac{D_{4}}{D_{1}D_{2}D_{3}D_{5}D_{6}D_{7}},\quad\quad\;\;
\T_{44}=\intd\frac{1}{D_{1}D_{2}D_{3}D_{5}D_{6}^{2}D_{7}}, \nonumber\\
\T_{45}&=\intd\frac{1}{D_{1}D_{2}D_{3}D_{4}D_{6}D_{7}},\quad\quad\;\;
\T_{46}=\intd\frac{D_{5}}{D_{1}D_{2}D_{3}D_{4}D_{6}D_{7}}, \nonumber\\
\T_{47}&=\intd\frac{1}{D_{1}D_{2}D_{3}D_{4}D_{6}D_{7}^{2}},\quad\quad\;\,
\T_{48}=\intd\frac{1}{D_{1}D_{2}D_{3}D_{4}D_{5}D_{6}D_{7}} .
\end{align}

\section{The Canonical Basis}
\label{app:CanBasis}

\begin{eqnarray}
f_{1}&=&\epsilon^2 \, p_4^2\, \mathcal{T}_{1} \, , \nonumber\\
f_{2}&=&\epsilon^2 \,t \,\mathcal{T}_{2} \, , \nonumber\\
f_{3}&=&\epsilon^2(m_B^2-p_4^2)\mathcal{T}_{3}\,+\,2m_B^2\,\epsilon^2\,\mathcal{T}_{4} \, , \nonumber\\
f_{4}&=&\epsilon^2 \,p_4^2\,\mathcal{T}_{4} \, , \nonumber\\
f_{5}&=&-2m_B^2\,\epsilon^2\mathcal{T}_{6}+\epsilon^2(t-m_B^2)\mathcal{T}_{5} \, , \nonumber\\
f_{6}&=&\epsilon^{2}\,t\,\mathcal{T}_{6} \, , \nonumber\\
f_{7}&=&\epsilon^{2}\,s\,\mathcal{T}_{7} \, , \nonumber\\
f_{8}&=&\epsilon^{2}(\epsilon-1)\,\mathcal{T}_{8} \, , \nonumber\\
f_{9}&=&\epsilon^{2}(s-m_B^2)\mathcal{T}_{9}+\epsilon^{2}(s-2m_B^2)\mathcal{T}_{10} \, , \nonumber\\
f_{10}&=&\epsilon^{2}\,s\,\mathcal{T}_{10} \, , \nonumber\\
f_{11}&=&\epsilon^{2}\,\sqrt{(p_{4}^2)^3}\sqrt{4m_B^2-p_4^2}\mathcal{T}_{11} \, , \nonumber\\
f_{12}&=&\epsilon^{2}\sqrt{p_4^2}\sqrt{4m_B^2-p_4^2}\,t\,\mathcal{T}_{12} \, , \nonumber\\
f_{13}&=&\epsilon^{2}\sqrt{p_4^2}\sqrt{4m_B^2-p_4^2}\,m_B^2\,\mathcal{T}_{13}\,
+\,\epsilon^{2}(m_B^2-p_4^2)\frac{\sqrt{4m_B^2-p_4^2}}{4\sqrt{p_4^2}}\,\mathcal{T}_{3}\nn\nonumber\\
&&+\,\epsilon^{2}(m_B^2-p_4^2)\frac{4m_B^2-p_4^2}{4\sqrt{p_4^2}}\,\mathcal{T}_{4}
-\epsilon^{2}(\epsilon-1)\frac{\sqrt{4m_B^2-p_4^2}}{2\sqrt{p_4^2}}\,\mathcal{T}_{8} \, , \nonumber\\
f_{14}&=&\epsilon^{2}\sqrt{4m_B^2-p_4^2}\sqrt{p_4^2}\,s\,\mathcal{T}_{14} \, , \nonumber\\
f_{15}&=&\epsilon^{3}(s-p_4^2)\mathcal{T}_{15} \, , \nonumber\\
f_{16}&=&\epsilon^{2}\frac{\sqrt{4m_B^2-p_4^2}\,\sqrt{p_4^2}\,(m_B^2\,p_4^2\,
-\,p_4^2\,s\,+\,s^{2})}{2s-p_4^2}\,\mathcal{T}_{16}\,
+\,\epsilon^{2}(\epsilon-1)\frac{\sqrt{4m_B^2-p_4^2}\,\sqrt{p_4^2}}{2p_4^2-4s}\,\mathcal{T}_{8}\nn\nonumber\\
&&+\,\frac{\sqrt{4m_B^2-p_4^2}\,\sqrt{p_4^2}(s-m_B^2)}{2(2p_4^2-4s)}\,\mathcal{T}_{9}
-3\epsilon^{3}\frac{\sqrt{4m_B^2-p_4^2}\,\sqrt{p_4^2}(s-p_4^2)}{2p_4^2-4s}\,\mathcal{T}_{15} \, , \nonumber\\
f_{17}&=&\epsilon^{3}(t-p_4^2)\,\mathcal{T}_{17} \, , \nonumber\\
f_{18}&=&\epsilon^{2}\frac{\sqrt{4m_B^2-p_4^2}\sqrt{p_4^2}\,(m_B^2\,p_4^2-p_4^2\,t\,+t^2)}{-p_4^2+2t}\,
\mathcal{T}_{18}\,-\,3\epsilon^{3}\frac{\sqrt{4m_B^2-p_4^2}\,\sqrt{p_4^2}(t-p_4^2)}{2p_4^2-4t}
\mathcal{T}_{17} \nn\nonumber\\
&&+\,\epsilon^{2}(\epsilon-1)\frac{\sqrt{4m_B^2-p_4^2}\,\sqrt{p_4^2}}{2p_4^2-4t}\,\mathcal{T}_{8}
 +\,\epsilon^{2}\frac{\sqrt{4m_B^2-p_4^2}\,\sqrt{p_4^2}(t-m_B^2)}{2(2p_4^2-4t)}\,\mathcal{T}_{5} \nn\nonumber\\
&&+\,\epsilon^{2}\frac{\sqrt{4m_B^2-p_4^2}\,\sqrt{p_4^2}\,t(t-m_B^2)}{2(p_4^2\,t-t^2)}
\,\mathcal{T}_{6} \, , \nonumber\\
f_{19}&=&\epsilon^{3}(t-p_4^2)\mathcal{T}_{19} \, , \nonumber\\
f_{20}&=&\epsilon^{2}\,\frac{m_B^2(t-p_4^2)(m_B^2-p_4^2+t)}{-2m_B^2-p_4^2+t}
\,\mathcal{T}_{20}\,+\,\epsilon^{3}\frac{(p_4^2-t)(-4m_B^2+p_4^2-t)}{2m_B^2+p_4^2-t}
\,\mathcal{T}_{19} \nn\nonumber\\
&& +\epsilon^{2}\frac{(p_4^2-t)t}{2m_B^2+p_4^2-t}\,\mathcal{T}_{2}
-\,\epsilon^{2}\frac{2m_B^2(m_B^2-p_4^2)}{-2m_B^2-p_4^2+t}\,\mathcal{T}_{3}\,
-\,\epsilon^{2}\frac{2m_B^2(2m_B^2+p_4^2)}{-2m_B^2-p_4^2+t}\,\mathcal{T}_{4} \, , \nonumber\\
f_{21}&=&\epsilon^{3}(s-p_4^2)\mathcal{T}_{21} \, , \nonumber\\
f_{22}&=&\frac{\epsilon^{2}\,m_B^2(s-p_4^2)(m_B^2-p_4^2+s)}{-2 m_B^2-p_4^2+s}\mathcal{T}_{22}+\frac{\epsilon^3 (p_4^2-s) (-4 m_B^2+p_4^2-s)}{2 m_B^2+p_4^2-s}\,\mathcal{T}_{21} \nn\nonumber\\
&& -\frac{2\epsilon ^2 m_B^2  (m_B^2-p_4^2)}{-2m_B^2-p_4^2+s}\,\mathcal{T}_{3}
-\frac{2\epsilon^{2}\,m_B^2(2m_B^2+p_4^2)}{-2m_B^2-p_4^2+s}\mathcal{T}_{4}\,
+\,\frac{\,2\epsilon^{2} m_B^2\,s}{-2m_B^2-p_4^2+s}\mathcal{T}_{7} \, , \nonumber\\
f_{23}&=&\epsilon^{2}(-1+2\epsilon)m_B^2\,p_4^2\,\mathcal{T}_{23}\,
+\,\epsilon^{2}p_4^4\mathcal{T}_{11}\,-\,\epsilon^{2}m_B^2\,p_4^2\,\mathcal{T}_{13} \, , \nonumber\\
f_{24}&=&\epsilon^{4}(t-p_4^2)\,\mathcal{T}_{24} \, ,\nonumber\\
f_{25}&=&\epsilon^{3}\,s\,t\,\mathcal{T}_{25} \, , \nonumber\\
f_{26}&=&\epsilon^{4}(s-p_4^2)\,\mathcal{T}_{26} \, , \nonumber\\
f_{27}&=&\epsilon^{4}(t-p_4^2)\,\mathcal{T}_{27} \, , \nonumber\\
f_{28}&=&\epsilon^{3}\,m_B^2\,(t-p_4^2)\mathcal{T}_{28} \, , \nonumber\\
f_{29}&=&\epsilon^{2}\,m_B^2\,p_4^2\,t\,\mathcal{T}_{29}\,
+\,2\,\epsilon^{4}\,t\,\mathcal{T}_{27}\,+\,\frac{1}{2}m_B^2\epsilon^{3}(p_4^2-t)\,\mathcal{T}_{28}\,
+\frac{1}{2}p_4^2\epsilon^{3}(p_4^2+t)\,\mathcal{T}_{30} \nn\nonumber\\
&& + \frac{\epsilon^{2}(-4m_B^2+p_4^2)t}{2m_B^2+p_4^2-t}\,\mathcal{T}_{2}
-\frac{\epsilon^{2}(4m_B^4-5m_B^2 p_4^2+p_4^4)}{-2m_B^2-p_4^2+t}\mathcal{T}_{3}\,
-\,\frac{\epsilon^ {2}(8m_B^4+2m_B^2 p_4^2-p_4^4)}{-2m_B^2-p_4^2+t}\,\mathcal{T}_{4} \nn\nonumber\\
&& + \frac{\epsilon^{2}(m_B^2-t)t}{2(p_4^2-2t)}\mathcal{T}_{5}\,
+\,\frac{ \epsilon ^2\,t (m_B^2-p_4^2+t)}{p_4^2-2 t}\T_{6}
-\frac{(\epsilon-1) \epsilon^2 (p_4^2-t)}{p_4^2-2 t}\mathcal{T}_{8}\,+\,p_4^2\,t\, \epsilon^{2}\,\mathcal{T}_{12} \nn\nonumber\\
&& + m_B^2\,p_4^2\, \epsilon ^2\,\T_{13}\,-\frac{\epsilon ^3 \left(3 p_4^4-5 p_4^2 t+t^2\right)}{p_4^2-2 t}\T_{17} +\frac{p_4^2 \epsilon ^2 (m_B^2 p_4^2+t (t-p_4^2))}{p_4^2-2 t}\T_{18} \nn\nonumber\\
&& -\frac{2 \epsilon ^3 (3 p_4^2-2 t) (m_B^2-p_4^2+t)}{2 m_B^2+p_4^2-t}\T_{19}\,
+\,\frac{\epsilon ^2\,m_B^2 (3 p_4^2-2 t) (m_B^2-p_4^2+t)}{2 m_B^2+p_4^2-t}\T_{20} \, , \nonumber\\
f_{30}&=& \epsilon ^3\,\sqrt{p_4^2}\,\sqrt{4 m_B^2-p_4^2}\,(t-p_4^2)\,\T_{30} \, , \nonumber\\
f_{31}&=&\epsilon ^4\,(s-p_4^2)\,\T_{31} \, , \nonumber\\
f_{32}&=&\epsilon ^3\,s (s-p_4^2)\,T_{32} \, , \nonumber\\
f_{33}&=&\epsilon^{2}\,m_B^2\,p_4^2\,s\,\T_{33}\,
+\,2 \epsilon ^4 (p_4^2+s)\,\T_{31}\,+\, \epsilon ^3\,s (s-p_4^2)\,\T_{32}\,
+\,\epsilon ^3\,p_4^2 (p_4^2+s)\,\T_{34} \nn\nonumber\\
&& + \frac{2 \epsilon ^2\,p_4^2 (p_4^2-m_B^2)}{2 m_B^2+p_4^2-s}\,\T_{3}
-\frac{2\epsilon ^2 \,p_4^2\, s }{2 m_B^2+p_4^2-s}\,\T_{4}\,
+\,\frac{2 \epsilon ^2\,s (4 m_B^2+3 p_4^2-2 s)}{2 m_B^2+p_4^2-s}\,\T_{7} \nn\nonumber\\
&& -\frac{p_4^2 (\epsilon -1) \epsilon ^2}{p_4^2-2 s}\,\T_{8}\,
+\,\frac{\epsilon ^2\,p_4^2 (m_B^2-s)}{2 (p_4^2-2 s)}\,\T_{9}
+\frac{ \epsilon ^2\,p_4^2 (m_B^2-s)}{p_4^2-2 s}\,\T_{10}\,
+\,2 \epsilon ^2\, m_B^2\, p_4^2\,\T_{13} \nn\nonumber\\
&& + 2 \epsilon ^2\,p_4^2\,s\,\T_{14}\,
+\,\frac{ \epsilon ^3\,p_4^2 (7 s-5 p_4^2)}{p_4^2-2 s}\,\T_{15}
+\frac{2 \epsilon ^2\,p_4^2\,(m_B^2 p_4^2+s (s-p_4^2))}{p_4^2-2 s}\,\T_{16} \nn\nonumber\\
&& - \frac{4 \epsilon ^3\,p_4^2 (m_B^2-p_4^2+s)}{2 m_B^2+p_4^2-s}\,\T_{21}\,
+\,\frac{2 m_B^2\, p_4^2\, \epsilon^{2}(m_B^2-p_4^2+s)}{2 m_B^2+p_4^2-s}\,\T_{22} \, , \nonumber\\
f_{34}&=&\epsilon^3 \,\sqrt{p_4^2} \sqrt{4 m_B^2-p_4^2} (s-p_4^2)\,\T_{34} \, , \nonumber\\
f_{35}&=&\epsilon ^4\,(s+t)\,\T_{35} \, , \nonumber\\
f_{36}&=&\epsilon ^3\, (-(m_B^2 (s+t)-s t))\,\T_{36}\,+\,2 \epsilon ^4 (s+t)\,\T_{35} \, , \nonumber\\
f_{37}&=&\epsilon ^3\,t (s-m_B^2)\,\T_{37}\,+\, \epsilon ^3 (s-m_B^2)t\,\T_{38}\, , \nonumber\\
f_{38}&=&\epsilon ^3\,s\,t\,\T_{38} \, , \nonumber\\
f_{39}&=&\epsilon ^3\,s\, (t-m_B^2)\,\T_{39}\,+\, \epsilon ^3\,s\, (t-m_B^2)\,\T_{40} \, , \nonumber\\
f_{40}&=& \epsilon ^3\,s\,t\,\mathcal{T}_{40} \, , \nonumber\\
f_{41}&=&\epsilon ^3 \sqrt{p_4^2}\, \sqrt{4 m_B^2-p_4^2}\,s\, t\,\T_{41} \, , \nonumber\\
f_{42}&=&\epsilon ^4\,t\, (s-p_4^2)\,\T_{42} \, , \nonumber\\
f_{43}&=&\epsilon ^4\,t\,\T_{43}\,+\,\epsilon ^4
\,p_4^2\,t\,\T_{42}\,-\epsilon ^4\,t\,\T_{35} \, , \nonumber\\
f_{44}&=&-\frac{\epsilon ^3\,\sqrt{p_4^2}\, \sqrt{4 m_B^2-p_4^2}\,t\, (m_B^2 p_4^2+s (s-p_4^2))}{p_4^2-2 s}\,\T_{44}\,
+\,\frac{\epsilon ^4\,\sqrt{p_4^2}\, \sqrt{4 m_B^2-p_4^2}\,t}{p_4^2-2 s}\,\T_{43} \nn\nonumber\\
&& + \frac{\epsilon ^4 \,\sqrt{p_4^2}\,\sqrt{4 m_B^2-p_4^2}\,(2 p_4^2-s)\,t}{p_4^2-2 s}\,\T_{42}\,
+ \frac{\epsilon^{2}\,(m_B^2-3p_4^2)\,\sqrt{p_4^2}\,\sqrt{4m_B^2-p_4^2}\,t}{3(p_4^2-2s)(2m_B^2+p_4^2-t)}\,
\T_{2} \nn\nonumber\\
&& - \frac{\epsilon ^2\,(m_B^2-p_4^2)\,\sqrt{p_4^2}\,\sqrt{4 m_B^2-p_4^2}\,(10 m_B^2-3 (3 p_4^2+t))}{12 (p_4^2-2 s) (2 m_B^2+p_4^2-t)}\,\T_{3} \nn\nonumber\\
&& + \frac{\epsilon ^2\,\sqrt{p_4^2}\, \sqrt{4 m_B^2-p_4^2} \left(-10 m_B^4+m_B^2 (p_4^2+3 t)+3 p_4^2 (p_4^2+t)\right)}{6 (p_4^2-2s) (2 m_B^2+p_4^2-t)}\,\T_{4} \nn\nonumber\\
&& + \frac{ \epsilon ^2\,\sqrt{p_4^2}\,\sqrt{4 m_B^2-p_4^2} (t-m_B^2) (3 p_4^2-2 t)}{8 (p_4^2-2 s) (p_4^2-2 t)}\,\T_{5} \nn\nonumber\\
&& + \frac{\epsilon ^2 \sqrt{p_4^2}\,\sqrt{4 m_B^2-p_4^2}\,(2 t (m_B^2+p_4^2)-3 m_B^2 p_4^2)}{4 
(p_4^2-2 s) (p_4^2-2 t)}\,\T_{6} \nn\nonumber\\
&& + \frac{(\epsilon -1) \epsilon ^2 \sqrt{p_4^2} \, \sqrt{4 m_B^2-p_4^2}\,(11 p_4^2-16 t)}{6 
(p_4^2-2 s) (p_4^2-2 t)}\,\T_{8}\,+\, \frac{\epsilon ^2 \sqrt{p_4^2}\, \sqrt{4 m_B^2-p_4^2} 
(s-m_B^2)}{8 (p_4^2-2 s)}\,\T_{9} \nn\nonumber\\
&& - \frac{\epsilon ^2\,m_B^2\, \sqrt{p_4^2} \, \sqrt{4 m_B^2-p_4^2}}{4 p_4^2-8 s}\,\T_{10} 
-\frac{\epsilon ^2\,\sqrt{(p_{4}^{2})^3}\,  \sqrt{4 m_B^2-p_4^2}\,t}{p_4^2-2 s}\,\T_{12} \nn\nonumber\\
&& - \frac{\epsilon ^2\,m_B^2 \sqrt{(p_{4}^2)^3} \, \sqrt{4 m_B^2-p_4^2}}{p_4^2-2 s}\,\T_{13} 
+\,\frac{\epsilon ^3\,\sqrt{p_4^2} \, \sqrt{4 m_B^2-p_4^2}\,(s-p_4^2)}{2 p_4^2-4 s}\,\T_{15} \nn\nonumber\\
&& + \frac{\epsilon ^3\,\sqrt{p_4^2} \, \sqrt{4 m_B^2-p_4^2} \left(2 p_4^4-2 p_4^2 t-t^2\right)}{
(p_4^2-2 s) (p_4^2-2 t)}\,\T_{17} \nn\nonumber\\
&& -\frac{\epsilon ^2\,\sqrt{(p_{4}^2)^3} \, \sqrt{4 m_B^2-p_4^2}
\,(m_B^2 p_4^2+t (t-p_4^2))}{(p_4^2-2 s) (p_4^2-2 t)}\,\T_{18} \nn\nonumber\\
&& + \frac{\epsilon ^3\,\sqrt{p_4^2}\,\sqrt{4 m_B^2-p_4^2} (m_B^2 (3 p_4^2-t)+2 p_4^2 
(t-p_4^2))}{(p_4^2-2 s) (2 m_B^2+p_4^2-t)}\,\T_{19} \nn\nonumber\\
&& - \frac{\epsilon ^2\,m_B^2 \sqrt{p_4^2}\, \sqrt{4 m_B^2-p_4^2} (7 p_4^2-t) (m_B^2-p_4^2+t)}{6 
(p_4^2-2 s) (2 m_B^2+p_4^2-t)}\,\T_{20} \nn\nonumber\\
&& +\frac{\epsilon ^4\,\sqrt{p_4^2} \, \sqrt{4 m_B^2-p_4^2}\,(p_4^2-s)}{p_4^2-2 s}\,\T_{26} \nn\nonumber\\
&& - \frac{\epsilon ^4 \sqrt{p_4^2} \, \sqrt{4 m_B^2-p_4^2}\,(p_4^2+t)}{p_4^2-2 s}\,\T_{27}
-\frac{\epsilon ^2\,m_B^2 \,\sqrt{(p_{4}^2)^3}\, \sqrt{4 m_B^2-p_4^2}\,t}{p_4^2-2 s}\,\T_{29} \nn\nonumber\\
&& - \frac{\epsilon ^3\,\sqrt{(p_{4}^2)^3} \, \sqrt{4 m_B^2-p_4^2}\,(p_4^2+t)}{2 (p_4^2-2 s)}\,\T_{30}\,
+\,\frac{\epsilon ^4\,\sqrt{p_4^2} \, \sqrt{4 m_B^2-p_4^2}\, (3 s+t)}{2 (p_4^2-2 s)}\,\T_{35} \nn\nonumber\\
&& + \frac{\epsilon ^3\,\sqrt{p_4^2} \, \sqrt{4 m_B^2-p_4^2}\, (m_B^2 (s+t)-s t)}{4 (p_4^2-2 s)}\,\T_{36} \nn\nonumber\\
&& + \frac{\epsilon ^3\,\sqrt{p_4^2} \, \sqrt{4 m_B^2-p_4^2}\,t\, (s-m_B^2)}{6 (p_4^2-2 s)}\,\T_{37}\,
-\frac{\epsilon ^3\,\sqrt{p_4^2} \, \sqrt{4 m_B^2-p_4^2}\,(m_B^2-3 s)\,t}{6 (p_4^2-2 s)}\,\T_{38}
\, , \nonumber\\
f_{45}&=&\epsilon ^4\,s\, (t-p_4^2)\,\T_{45} \, , \nonumber\\
f_{46}&=&\epsilon ^4\,s\,\T_{46}\,+\, \epsilon ^4\,p_4^2\,s\,\T_{45}\,-\epsilon ^4\,s\,\T_{35} \, , \nonumber\\
f_{47}&=&-\frac{\epsilon ^3\,\sqrt{p_4^2} \sqrt{4 m_B^2-p_4^2}\,s\, (m_B^2 p_4^2+t (t-p_4^2))}{p_4^2-2 t}\,\T_{47}\,+\,\frac{\epsilon ^4\,\sqrt{p_4^2} \sqrt{4 m_B^2-p_4^2}\,s}{p_4^2-2 t}\,\T_{46} \nn\nonumber\\
&& + \frac{\epsilon ^4\,\sqrt{p_4^2} \, \sqrt{4 m_B^2-p_4^2} (2 p_4^2-t)\,s}{p_4^2-2 t}\,\T_{45} \nn\nonumber\\
&& - \frac{\epsilon ^2\,\sqrt{p_4^2} \, \sqrt{4 m_B^2-p_4^2}(m_B^2-p_4^2)(10 m_B^2-3 (3 p_4^2+s))}{12 
(p_4^2-2 t) (2 m_B^2+p_4^2-s)}\,\T_{3} \nn\nonumber\\
&& + \,\frac{\epsilon ^2\,\sqrt{p_4^2} \, \sqrt{4 m_B^2-p_4^2} \left(-10 m_B^4+m_B^2 (p_4^2+3 s)+3 p_4^2 (p_4^2+s)\right)}{6 (p_4^2-2
   t) (2 m_B^2+p_4^2-s)}\,\T_{4} \nn\nonumber\\
&& + \frac{\epsilon ^2\,\sqrt{p_4^2}\,\sqrt{4 m_B^2-p_4^2}\, (t-m_B^2)}{8 (p_4^2-2 t)}\,\T_{5}\,-\frac{\epsilon ^2\,m_B^2\,\sqrt{p_4^2} \, \sqrt{4 m_B^2-p_4^2}}{4 p_4^2-8 t}\,\T_{6} \nn\nonumber\\
&& + \frac{\epsilon ^2\,\sqrt{p_4^2}\, \sqrt{4 m_B^2-p_4^2}\,s\,(m_B^2-3 p_4^2)}{3 (p_4^2-2 t) (2 m_B^2+p_4^2-s)}\,\T_{7}+\frac{(\epsilon -1) \epsilon ^2\,\sqrt{p_4^2}\, \sqrt{4 m_B^2-p_4^2} (11 p_4^2-16 s)}{6 (p_4^2-2 s) (p_4^2-2 t)}\,\T_{8} \nn\nonumber\\
&& + \frac{\epsilon ^2\,\sqrt{p_4^2} \, \sqrt{4 m_B^2-p_4^2}\, (s-m_B^2) (3 p_4^2-2 s)}{8 (p_4^2-2 s) (p_4^2-2 t)}\,\T_{9} \nn\nonumber\\
&&+\frac{\epsilon ^2\,\sqrt{p_4^2} \, \sqrt{4 m_B^2-p_4^2}\,(2 s (m_B^2+p_4^2)-3 m_B^2 p_4^2)}{4 (p_4^2-2 s) (p_4^2-2 t)}\,\T_{10}\,-\frac{\epsilon ^2\,m_B^2 \sqrt{p_4^2}^3 \, \sqrt{4 m_B^2-p_4^2}}{p_4^2-2 t}\,\T_{13} \nn\nonumber\\
&&-\frac{\epsilon ^2\,\sqrt{p_4^2}^3 \, \sqrt{4 m_B^2-p_4^2}\,s}{p_4^2-2 t}\,\T_{14}+\frac{\epsilon ^3\,\sqrt{p_4^2} \, \sqrt{4 m_B^2-p_4^2} \left(2 p_4^4-2 p_4^2 s-s^2\right)}{(p_4^2-2 s) (p_4^2-2 t)}\,\T_{15} \nn\nonumber\\
&&-\frac{\epsilon ^2\,\sqrt{p_4^2}^3 \, \sqrt{4 m_B^2-p_4^2}\, (m_B^2 p_4^2+s (s-p_4^2))}{(p_4^2-2 s) (p_4^2-2 t)}\,\T_{16}+\frac{\epsilon ^3\,\sqrt{p_4^2} \, \sqrt{4 m_B^2-p_4^2}\, (t-p_4^2)}{2 p_4^2-4 t}\,\T_{17} \nn\nonumber\\
&&+\,
\frac{ \epsilon ^3\,\sqrt{p_4^2}\, \sqrt{4 m_B^2-p_4^2} (m_B^2 (3 p_4^2-s)+2 p_4^2 (s-p_4^2))}{(p_4^2-2 t) (2 m_B^2+p_4^2-s)}\,\T_{21} \nn\nonumber\\
&&-\frac{\epsilon ^2\,m_B^2 \,\sqrt{p_4^2} \, \sqrt{4 m_B^2-p_4^2} (7 p_4^2-s) (m_B^2-p_4^2+s)}{6 (p_4^2-2 t) (2 m_B^2+p_4^2-s)}\,\T_{22} \nn\nonumber\\
&& + \frac{\epsilon ^4\,\sqrt{p_4^2} \, \sqrt{4 m_B^2-p_4^2} (p_4^2-t)}{p_4^2-2 t}\,\T_{24}\,
-\frac{\epsilon ^4\,\sqrt{p_4^2} \, \sqrt{4 m_B^2-p_4^2} (p_4^2+s)}{p_4^2-2 t}\,\T_{31} \nn\nonumber\\
&& -\frac{\epsilon ^2 \,m_B^2\,s\, \sqrt{p_4^2}^3 \,\sqrt{4 m_B^2-p_4^2}}{p_4^2-2 t}\,\T_{33}\,
-\frac{\epsilon ^3\,\sqrt{p_4^2}^3 \, \sqrt{4 m_B^2-p_4^2}\, (p_4^2+s)}{2 (p_4^2-2 t)}\,\T_{34} \nn\nonumber\\
&& + \frac{\epsilon ^4\,\sqrt{p_4^2} \, \sqrt{4 m_B^2-p_4^2}\, (s+3 t)}{2 (p_4^2-2 t)}\,\T_{35}+\frac{\epsilon ^3\,\sqrt{p_4^2} \, \sqrt{4 m_B^2-p_4^2}\,(m_B^2 (s+t)-s t)}{4 (p_4^2-2 t)}\,\T_{36} \nn\nonumber\\
&& + \frac{\epsilon ^3\,\sqrt{p_4^2} \, \sqrt{4 m_B^2-p_4^2}\,s\, (t-m_B^2)}{6 (p_4^2-2 t)}\,\T_{39}\,-\frac{\epsilon ^3\sqrt{p_4^2}\, \sqrt{4 m_B^2-p_4^2}\,s\, (m_B^2-3 t)}{6 (p_4^2-2 t)}\,\T_{40} \, , \nonumber\\
f_{48}&=&\epsilon ^4\,p_4^2\,s\,t\,\T_{48}\,+\,
\frac{\epsilon ^2 (m_B^2-3 p_4^2)\sqrt{p_4^2}\, \sqrt{4 m_B^2-p_4^2}t}{3 (p_4^2-2 s) (2 m_B^2+p_4^2-t)}\,\T_{2} + A(s,t,p_4^2)\mathcal{T}_3+B(s,t,p_4^2)\mathcal{T}_4\nn\nonumber\\
&&+\frac{ \epsilon ^2\sqrt{p_4^2}\, \sqrt{4 m_B^2-p_4^2} \,(m_B^2-t) (-2 p_4^2+s+t)}{4 (p_4^2-2 s) (p_4^2-2 t)}\,\T_{5} \nn\nonumber\\
&& + \frac{\epsilon^{2}\sqrt{p_4^2}\, \sqrt{4 m_B^2-p_4^2} (m_B^2 (-2 p_4^2+s+t)+p_4^2 t)}{2 (p_4^2-2 s) (p_4^2-2 t)}\,\T_{6} \nn\nonumber\\
&&+\frac{ (\epsilon -1) \epsilon ^2\sqrt{p_4^2}\, \sqrt{4 m_B^2-p_4^2} (11 p_4^2-8 (s+t))}{3 (p_4^2-2 s) (p_4^2-2 t)}\,\T_{8} \nn\nonumber\\
&&+\frac{\epsilon ^2\,\sqrt{p_4^2} \, \sqrt{4 m_B^2-p_4^2}\, (m_B^2-s) (-2 p_4^2+s+t)}{4 (p_4^2-2 s) (p_4^2-2 t)}\,\T_{9}\nn\nonumber\\
&& +\,\frac{\epsilon ^2\,\sqrt{p_4^2} \, \sqrt{4 m_B^2-p_4^2}\,(m_B^2 (-2 p_4^2+s+t)+p_4^2 s)}{2 (p_4^2-2 s) (p_4^2-2 t)}\,\T_{10}
-\frac{\epsilon ^2\,\sqrt{p_4^2}^3 \, \sqrt{4 m_B^2-p_4^2}\,t}{p_4^2-2 s}\,\T_{12} \nn\nonumber\\
&& +\,\frac{2\epsilon ^2\, m_B^2\, ^{3/2} \, \sqrt{4 m_B^2-p_4^2}\, (-p_4^2+s+t)}{(p_4^2-2 s) 
(p_4^2-2 t)}\,\T_{13}
-\frac{\epsilon ^2\,\sqrt{p_4^2}^3 \,  \sqrt{4 m_B^2-p_4^2}\,s}{p_4^2-2 t}\,\T_{14}\nn\nonumber\\
&& 
+\frac{\epsilon ^3\,\sqrt{p_4^2} \, \sqrt{4 m_B^2-p_4^2} \left(3 p_4^4+p_4^2 (2 t-3 s)-2 s (s+t)\right)}{2 (p_4^2-2s) (p_4^2-2 t)}\,\T_{15}\nn\nonumber\\
&& -\frac{\epsilon ^2\,\sqrt{p_4^2}^3\, \sqrt{4 m_B^2-p_4^2}\, (m_B^2 p_4^2+s (s-p_4^2))}{(p_4^2-2 s) (p_4^2-2 t)}\,\T_{16} \nn\nonumber\\
&&+\frac{\epsilon ^3\,\sqrt{p_4^2} \, \sqrt{4 m_B^2-p_4^2}\,\left(3 p_4^4+p_4^2 (2 s-3 t)-2 t (s+t)\right)}{2 (p_4^2-2s) (p_4^2-2 t)}\,\T_{17}\nn\nonumber\\
&&-\frac{\epsilon ^2\,\sqrt{p_4^2}^3 \, \sqrt{4 m_B^2-p_4^2}\,(m_B^2 p_4^2+t (t-p_4^2))}{(p_4^2-2 s) (p_4^2-2 t)}\,\T_{18}\nn\nonumber\\
&&+\frac{\epsilon ^3\,\sqrt{p_4^2} \, \sqrt{4 m_B^2-p_4^2}\,(m_B^2 (3 p_4^2-t)+2 p_4^2 (t-p_4^2))}{(p_4^2-2 s) (2 m_B^2+p_4^2-t)}\,\T_{19}\nn\nonumber\\
&&-\frac{\epsilon ^2\,m_B^2\, \sqrt{p_4^2}\, \sqrt{4 m_B^2-p_4^2} (7 p_4^2-t) (m_B^2-p_4^2+t)}{6 (p_4^2-2 s) (2 m_B^2+p_4^2-t)}\,\T_{20} \nn\nonumber\\
&&+\frac{\epsilon ^3\,\sqrt{p_4^2} \,\sqrt{4 m_B^2-p_4^2}\, (m_B^2 (3 p_4^2-s)+2 p_4^2 (s-p_4^2))}{(p_4^2-2 t) (2 m_B^2+p_4^2-s)}\,\T_{21}\nn\nonumber\\
&&-\frac{\epsilon ^2 \,m_B^2\, \sqrt{p_4^2} \,\sqrt{4 m_B^2-p_4^2}\,(7 p_4^2-s) (m_B^2-p_4^2+s)}{6 (p_4^2-2 t) (2 m_B^2+p_4^2-s)}\,\T_{22} \nn\nonumber\\
&&+\frac{\epsilon ^4\,\sqrt{p_4^2} \, \sqrt{4 m_B^2-p_4^2}\, (p_4^2-t)}{p_4^2-2 t}\,\T_{24}\,+\,
\frac{\epsilon ^4\,\sqrt{p_4^2} \, \sqrt{4 m_B^2-p_4^2}\, (p_4^2-s)}{p_4^2-2 s}\,\T_{26} \nn\nonumber\\
&&-\frac{\epsilon ^4\,\sqrt{p_4^2} \, \sqrt{4 m_B^2-p_4^2}\,(p_4^2+t)}{p_4^2-2 s}\,\T_{27} 
+ \epsilon ^3\,m_B^2 \, (t-p_4^2)\,\T_{28}\, \nn\nonumber\\
&& -\frac{\epsilon ^2\,m_B^2\, \sqrt{p_4^2}^3 \, \sqrt{4 m_B^2-p_4^2}\,t}{p_4^2-2 s}\,\T_{29}\,
-\frac{\epsilon ^3\,\sqrt{p_4^2}^3 \, \sqrt{4 m_B^2-p_4^2}\, (p_4^2+t)}{2 (p_4^2-2 s)}\,\T_{30} \nn\nonumber\\
&&-\frac{\epsilon ^4\,\sqrt{p_4^2} \, \sqrt{4 m_B^2-p_4^2}\,(p_4^2+s)}{p_4^2-2 t}\,\T_{31}\,+\,\epsilon ^3\,s\, (s-p_4^2)\,\T_{32}\,\nn\nonumber\\
&&
-\frac{\epsilon ^2\,m_B^2\, \sqrt{p_4^2}^3 \, \sqrt{4 m_B^2-p_4^2}\,s}{p_4^2-2 t}\,\T_{33}
-\frac{\epsilon ^3\,\sqrt{p_4^2}^3\, \sqrt{4 m_B^2-p_4^2}\,(p_4^2+s)}{2 (p_4^2-2 t)}\,\T_{34} \nn\nonumber\\
&& + \frac{\epsilon ^4\,\sqrt{p_4^2} \,\sqrt{4 m_B^2-p_4^2} \left(2 p_4^2 (s+t)-s^2-6 s t-t^2\right)}{(p_4^2-2 s)(p_4^2-2 t)}\,\T_{35}\nn\nonumber\\
&& + \frac{\epsilon ^3 \,\sqrt{p_4^2} \,\sqrt{4 m_B^2-p_4^2}\, (m_B^2 (s+t)-s t) (p_4^2-s-t)}{2 (p_4^2-2 s) (p_4^2-2 t)}\,\T_{36} \nn\nonumber\\
&&+\frac{\epsilon ^3\,\sqrt{p_4^2} \, \sqrt{4 m_B^2-p_4^2}\,t\, (s-m_B^2)}{6 (p_4^2-2 s)}\,\T_{37}\,
-\frac{\epsilon ^3\,\sqrt{p_4^2}\,  \sqrt{4 m_B^2-p_4^2} (m_B^2-3 s)\,t}{6 (p_4^2-2 s)}\,\T_{38}\nn\nonumber\\
&& + \frac{\epsilon ^3\,\sqrt{p_4^2}\,\sqrt{4 m_B^2-p_4^2}\,s (t-m_B^2)}{6 (p_4^2-2 t)}\,\T_{39}
-\frac{\epsilon ^3\,\sqrt{p_4^2}\,\sqrt{4 m_B^2-p_4^2}\,s\, (m_B^2-3 t)}{6 (p_4^2-2 t)}\,\T_{40}\nn\nonumber\\
&& + \frac{\epsilon ^4\,\sqrt{p_4^2}\, t \left(-s \sqrt{4 m_B^2-p_4^2}+2 p_4^2 \sqrt{4 m_B^2-p_4^2}+\sqrt{p_4^2} (p_4^2-2
   s)\right)}{p_4^2-2 s}\,\T_{42} \nn\nonumber\\
&&+\frac{\epsilon ^4\,\sqrt{p_4^2} \, \sqrt{4 m_B^2-p_4^2}\,t}{p_4^2-2 s}\,\T_{43}\,
-\frac{\epsilon ^3\,\sqrt{p_4^2}\,\sqrt{4 m_B^2-p_4^2}\,t\, (m_B^2 p_4^2+s (s-p_4^2))}{p_4^2-2 s}\,\T_{44} \nn\nonumber\\
&&+\frac{\epsilon ^4\,s\,\sqrt{p_4^2}\left(-t \sqrt{4 m_B^2-p_4^2}+2 p_4^2 \sqrt{4 m_B^2-p_4^2}+\sqrt{p_4^2} (p_4^2-2t)\right)}{p_4^2-2 t}\,\T_{45}\nn\nonumber\\
&& + \frac{\epsilon ^4\,s\,\sqrt{p_4^2}\,\sqrt{4 m_B^2-p_4^2}}{p_4^2-2 t}\,\T_{46}
-\frac{\epsilon ^3\,s\,\sqrt{p_4^2} \, \sqrt{4 m_B^2-p_4^2} (m_B^2 p_4^2+t (t-p_4^2))}{p_4^2-2 t}\,\T_{47} \, ,
\end{eqnarray} 

where
\bea
A(s,t,p_4^2) & = & -\frac{(m_w^2-p_4^2) \sqrt{p_4^2}\sqrt{4 m_B^2-p_4^2} \left(20 m_B^4 (p_4^2-s-t)+m_B^2 \left(2 (3 s+t) (s+3 t)-8 p_4^4\right)\right)}{6 (p_4^2-2 s) (p_4^2-2 t) (2 m_B^2+p_4^2-s) (2 m_B^2+p_4^2-t)}\nn \nonumber\\
&& + \frac{(m_w^2-p_4^2) \sqrt{p_4^2}\sqrt{4 m_B^2-p_4^2} \left(3 p_4^6-4 p_4^4 (s+t)-p_4^2 \left(s^2-5 s t+t^2\right)+s t (s+t)\right)}{2 (p_4^2-2 s) (p_4^2-2 t) (2 m_B^2+p_4^2-s) (2 m_B^2+p_4^2-t)} \nonumber\\
B(s,t,p_4^2) & = & -\frac{\sqrt{p_4^2} \sqrt{4 m_B^2-p_4^2} (2 m_B^2+p_4^2) s \left(-10 m_B^4-3 m_B^2 p_4^2+3 p_4^4\right)}{3 (p_4^2-2 s) (p_4^2-2 t) (2 m_B^2+p_4^2-s) (2 m_B^2+p_4^2-t)} \nn \nonumber\\
&& -\frac{\sqrt{p_4^2} \sqrt{4 m_B^2-p_4^2} (2 m_B^2+p_4^2) \left(3 s^2 (m_B^2+p_4^2)-p_4^2 (2 m_B^2+p_4^2) (3 p_4^2-5 m_B^2)\right)}{3 (p_4^2-2 s) (p_4^2-2 t) (2 m_B^2+p_4^2-s) (2 m_B^2+p_4^2-t)} \nn \nonumber\\
&& -\frac{t^2\sqrt{p_4^2} \sqrt{4 m_B^2-p_4^2} (m_B^2+p_4^2)}{(p_4^2-2 s) (p_4^2-2 t) (2 m_B^2+p_4^2-t)} \nn \nonumber\\
&& + \frac{t\sqrt{p_4^2}\sqrt{4 m_B^2-p_4^2} (2 m_B^2+p_4^2) \left(10 m_B^4+3 m_B^2 p_4^2-3 p_4^4\right)}{3 (p_4^2-2 s) (p_4^2-2 t) (2 m_B^2+p_4^2-s) (2 m_B^2+p_4^2-t)} \nn \nonumber\\
&& -\frac{st\sqrt{p_4^2}\sqrt{4 m_B^2-p_4^2} \left(20 m_B^4+m_B^2 (p_4^2-3 s)-3 p_4^2 (p_4^2+s)\right)}{3 (p_4^2-2 s) (p_4^2-2 t) (2 m_B^2+p_4^2-s) (2 m_B^2+p_4^2-t)}\,.
\eea

\bibliographystyle{JHEP}
\bibliography{biblio}

\end{document}